\documentclass[app,amsmath,amssymb,reprint,superscriptaddress,floatfix]{revtex4-1}

\usepackage{graphicx}
\usepackage{dcolumn}
\usepackage{bm}

\usepackage[utf8]{inputenc}
\usepackage[T1]{fontenc}
\usepackage{mathptmx}
\usepackage{siunitx}
\usepackage{hyperref}
\usepackage{float}
\usepackage{graphicx}
\usepackage{dcolumn}
\usepackage{bm}

\renewcommand{\copyright}{\textsuperscript{\tiny{\textcopyright}}}

\begin{document}


\title{Accurate modeling and characterization of photothermal forces in optomechanics}

\author{Andr\'{e} G. Primo}
\altaffiliation{These authors contributed equally for this work.}
\affiliation{Applied Physics Department and Photonics Research Center, ``Gleb Wataghin'' Institute of Physics, University of Campinas, 13083-859, Campinas, SP, Brazil}
\affiliation{Photonics Research Center, University of Campinas, 13083-859, Campinas, SP, Brazil}

\author{Cau\^{e} M. Kersul}
\altaffiliation{These authors contributed equally for this work.}
\affiliation{Applied Physics Department and Photonics Research Center, ``Gleb Wataghin'' Institute of Physics, University of Campinas, 13083-859, Campinas, SP, Brazil}
\affiliation{Photonics Research Center, University of Campinas, 13083-859, Campinas, SP, Brazil}

\author{Rodrigo Benevides}
\affiliation{Applied Physics Department and Photonics Research Center, ``Gleb Wataghin'' Institute of Physics, University of Campinas, 13083-859, Campinas, SP, Brazil}
\affiliation{Photonics Research Center, University of Campinas, 13083-859, Campinas, SP, Brazil}

\author{Nat\'{a}lia C. Carvalho}
\affiliation{Applied Physics Department and Photonics Research Center, ``Gleb Wataghin'' Institute of Physics, University of Campinas, 13083-859, Campinas, SP, Brazil}
\affiliation{Photonics Research Center, University of Campinas, 13083-859, Campinas, SP, Brazil}

\author{Micha\"{e}l M\'{e}nard}
\affiliation{Department of Computer Science, Universit\'{e} du Qu\'{e}bec \`{a} Montr\'{e}al, Montr\'{e}al, Canada}

\author{Newton C. Frateschi}
\affiliation{Applied Physics Department and Photonics Research Center, ``Gleb Wataghin'' Institute of Physics, University of Campinas, 13083-859, Campinas, SP, Brazil}
\affiliation{Photonics Research Center, University of Campinas, 13083-859, Campinas, SP, Brazil}

\author{Pierre-Louis de Assis}
\affiliation{Applied Physics Department and Photonics Research Center, ``Gleb Wataghin'' Institute of Physics, University of Campinas, 13083-859, Campinas, SP, Brazil}

\author{Gustavo S. Wiederhecker}
\affiliation{Applied Physics Department and Photonics Research Center, ``Gleb Wataghin'' Institute of Physics, University of Campinas, 13083-859, Campinas, SP, Brazil}
\affiliation{Photonics Research Center, University of Campinas, 13083-859, Campinas, SP, Brazil}

\author{Thiago P. Mayer Alegre}
\email{alegre@unicamp.br}
\affiliation{Applied Physics Department and Photonics Research Center, ``Gleb Wataghin'' Institute of Physics, University of Campinas, 13083-859, Campinas, SP, Brazil}
\affiliation{Photonics Research Center, University of Campinas, 13083-859, Campinas, SP, Brazil}

\begin{abstract}
Photothermal effects have been pointed out as prominent sources of forces in optomechanical systems, competing with the standard radiation pressure interactions. In this Article, we derive a novel and accurate model for the prediction of photothermal forces and establish how some previous proposals can be complemented to yield precise results. As a proof-of-concept, we perform numerical and experimental tests on GaAs microdisks cavities and obtain striking agreement with our framework, revealing the importance of considering surface photothermal forces and the effects of multiple thermal modes in microphotonic devices.
\end{abstract}

\maketitle

\section*{Introduction} Microscale photonic devices revolutionized the field of cavity optomechanics over the past two decades. The ability to selectively control the photon-phonon interaction through the detuning between optical resonances and external laser sources led to novel applications, ranging from nonlinear dynamics~\cite{Aspelmeyer2014CavityOptomechanics, Lemonde2016EnhancedAmplification} to the quantum manipulation of mechanical degrees of freedom~\cite{Chan2011, Safavi-Naeini2012, Forsch2020}. The extreme confinement of the optical fields and small effective masses result in devices with enhanced optomechanical effects, which may arise from distinct and competing mechanisms, such as photothermal (bolometric) forces~\cite{Woolf2013OptomechanicalMembranes}, radiation pressure~\cite{Wiederhecker2019BrillouinStructures}, and piezoelectricity~\cite{Carvalho2019Piezo-optomechanicalResonator}. While the last two are built upon a robust theoretical framework, based on optical and mechanical modal analysis, photothermal forces in optomechanical systems are often treated with phenomenological models that require a complete experimental characterization of the structures as input~\cite{Woolf2013OptomechanicalMembranes, Metzger2008Self-InducedBackaction,Barton}. The absence of an accurate and predictive photothermal force model has hindered its understanding and control within photonic devices. 

\begin{figure}[ht!]
\includegraphics[width = 8.5cm]{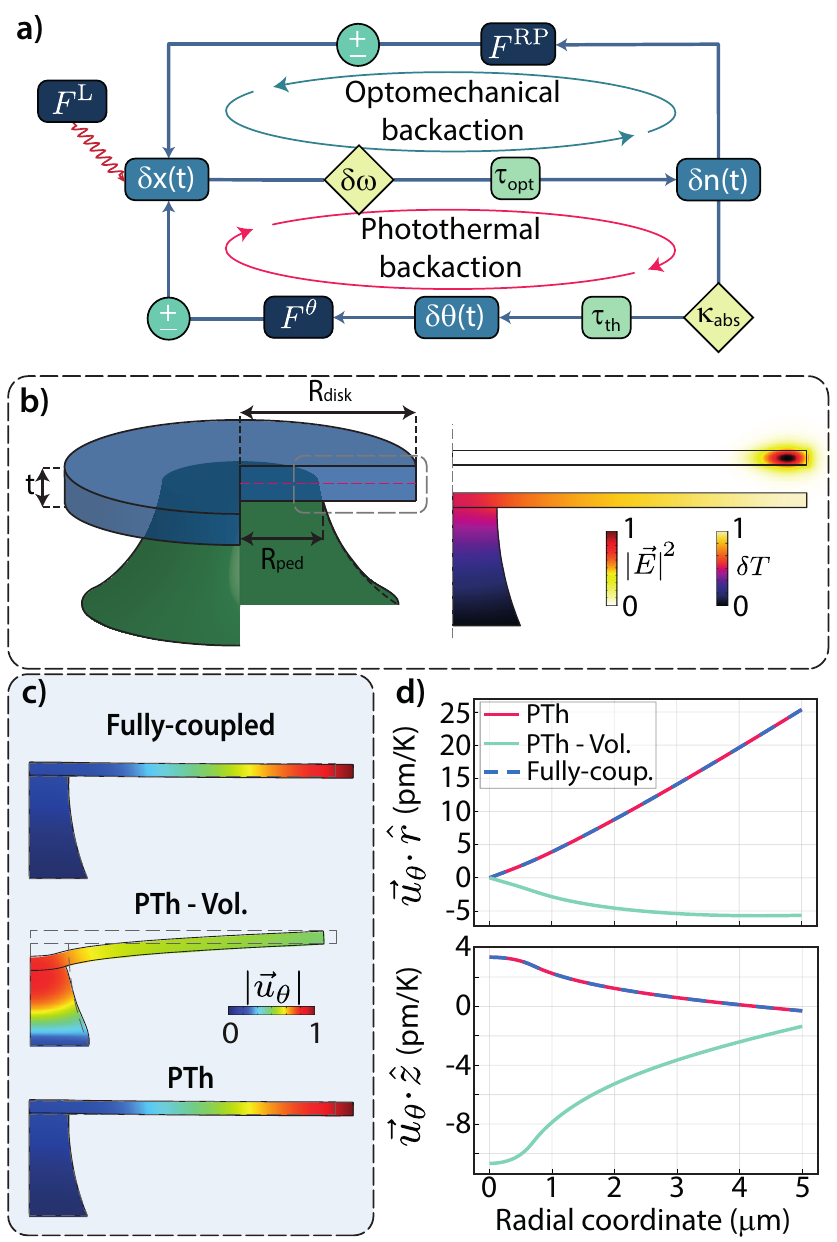}
\caption{\small{\textbf{a)} Photothermal and optomechanical backaction cycles. A Lagevin-type force $F^\text{L}$ induces fluctuations, $\delta x$, in the mechanical position. As a consequence, the optical resonance frequency is shifted by $\delta \omega$ causing a delayed modulation in the number of circulating photons $\delta n$. If optomechanical backaction is considered, optics and mechanics are coupled through a force $F^\text{RP}$, caused by radiation pressure and electrostriction. On the other hand, PTh backaction is obtained if absorption occurs, causing fluctuations in temperature $\delta \theta$. Thermal stresses modeled by the PTh force $F^\theta$ act back on the mechanics inducing cooling or amplification ($\pm$) of mechanical modes.  \textbf{b)} Geometric parameters for microdisk resonators, along with optical mode and temperature profile induced by absorption. \textbf{c)} Thermally induced deformations on disk structure, calculate through a fully-coupled thermoelastic model, volume PTh, and full PTh forces. \textbf{d)} Displacement field on the $\hat r$ and $\hat z$ directions as a function of the radial coordinate. The projections were evaluated at half the thickness ($t/2$) of the GaAs membrane, depicted by the red dashed curve in  \textbf{b)}.}}
\label{fig1}
\end{figure}

Despite the high optical quality factors ($Q>10^4$) of typical optomechanical resonators, absorptive losses may significantly impact the dynamics of mechanical modes~\cite{Hauer2019DuelingCavity,Guha2017HighSemiconductors}. As depicted in Fig.~\ref{fig1} \textbf{a)}, Brownian noise-driven mechanical motion modulates the number of circulating photons in the cavity, which, in association with absorption, drives oscillations in the temperature of the system. Finally, thermally-induced stresses couple back to the mechanical domain and close a feedback loop that defines the so-called photothermal backaction~\cite{Pinard2008, DeLiberato2011QuantumCooling}. The finite thermal and optical response times yield forces that are time-delayed relative to mechanical oscillations, allowing for the cooling~\cite{Metzger2004CavityMicrolever,Usami2012OpticalNanomembrane} and amplification~\cite{Metzger2008Self-InducedBackaction,Barton} of mechanical normal modes in a range of dielectric and plasmonic~\cite{Zhu2016PlasmonicResonances} resonators. This extra degree of freedom opens up new possibilities, such as thermally-mediated optomechanical ground state cooling in the bad-cavity regime~\cite{DeLiberato2011QuantumCooling}.
   
In this work, we propose and demonstrate a model for photothermal (PTh) forces. We introduce a novel mathematical treatment for the description of the thermal fields that ultimately allows the prediction of the PTh response in devices with arbitrary geometries. It is built upon thermal modal analysis~\cite{Duwel2006EngineeringDamping} and perturbation theory under a linear diffusive heat transfer regime, overcoming the limitations of several previous models based on a phenomenological treatment of PTh effects. As an example, we perform experiments in a GaAs microdisk cavity that remarkably agree with our predictions.

\section*{Model} 
The mechanical system is described by the equation of motion: $\rho\ddot{\vec{U}} = \bm \nabla\cdot \boldsymbol \sigma$, where $\vec{U}$ denotes the displacement field, and $\boldsymbol \sigma$ the stress tensor. In thermo-elasticity, the self-consistency of this problem requires a constitutive relation linking the stress tensor to the displacement and thermal fields. Since the stress arises solely from elastic deformations~\cite{Landau1986TheoryElasticity}, it is necessary to split the strain of the system, $\mathbf S = \frac{1}{2}(\bm \nabla \vec{U}+\bm \nabla^T \vec{U}) = \hat {\bm\nabla} \vec{U}$, into elastic ($ \mathbf S^x $) and thermal ($ \mathbf S^\theta $) components as $\mathbf S = \mathbf S^\theta + \mathbf S^x$. The constitutive relation then reads: $\boldsymbol \sigma = \mathbf c \mkern1mu{:} \mathbf S^x = \mathbf c \mkern1mu{:} \mathbf S - \mathbf c \mkern1mu{:} \mathbf S^\theta$~\cite{auld}, where $\mathbf c$ is the stiffness tensor, and ``$\mkern1mu{:}$'' denotes the tensor contraction operation. Due to the thermal strain contribution, the free-boundary condition ($\boldsymbol \sigma \cdot \hat n = 0$), commonly used in micromechanical devices, leads to a temperature-dependent surface-traction on $\mathbf S$ that acts as a drive for the mechanical fields, as detailed in section S1 of the Supplemental Material. In some previous formulations of PTh forces in optomechanical systems, this subtlety has been neglected~\cite{Schliesser2010CavityMicro-Resonators} and can lead to inaccurate predictions of the PTh response of optomechanical resonators.

A simple way to account for the boundary conditions is by calculating the PTh force directly from the work done by thermally-induced stresses on a given mechanical mode~\cite{Duwel2006EngineeringDamping, Murthy2016, Perez2009}. This is done in a linear approximation, where the elastic strain  can be decomposed in the mechanical normal modes of the system ($\vec{u}_n$) as $\mathbf{S}^x = \sum_n x_n(t) \hat{\bm \nabla} \vec u_n(\vec{r})=\sum_n x_n(t) \mathbf{S}_n^x(\vec{r})$, uncoupling the photothermal forces acting on each of the modes. The $x_n(t)$ are the normal mode amplitudes and are analogous to the generalized coordinates in analytical mechanics~\cite{Goldstein2007}. The present calculation allows, to first order, direct access to an expression for the lumped PTh force on a mechanical mode~\cite{Primo2019ModellingOptomechanics,Primo2020ThermodynamicOptomechanics,QuantumPress,Hetnarski2019BasicThermoelasticity}:
\begin{equation}
    F^{\theta}_n(t) = \frac{\partial}{\partial x_n}\int {\mathbf{S}}^x\mkern1mu{:}(\mathbf{c}\mkern1mu{:}\mathbf{S}^\theta) dV=\int {\mathbf{S}_n^x}\mkern1mu{:}(\mathbf{c}\mkern1mu{:}\mathbf{S}^\theta) dV.
     \label{main}
\end{equation}
A similar reasoning and expression was successfully used to estimate thermoelastic damping in MEMS~\cite{Duwel2006EngineeringDamping}, and displayed exceptional agreement with experimental values.
Note that a partial differentiation in the amplitude $x_n(t)$ is performed, where the index $n$ denotes the mechanical mode in which we are evaluating the PTh force. The first integral in the above expressions resembles the known (elastic) strain energy~\cite{Cleland2013FoundationsApplications}. This association allows us to interpret it as the energy transferred between thermal and mechanical domains. 

The lumped photothermal force in Eq.~\ref{main} can be rewritten as the sum of a volume and a surface contributions, $ F^{\theta}_n = F^{\theta\text{-Vol.}}_n + F^{\theta\text{-Sur.}}_n $, where:
\begin{align}
        F^{\,\theta\text{-Vol.}}_n &= - \int \vec{u}_n \cdot \big( \bm\nabla \cdot (\mathbf{c}\mkern1mu{:}\mathbf{S}^\theta) \big) \, dV,\label{force_vol}\\
        F^{\theta\text{-Sur.}}_n &= \int \vec{u}_n \cdot (\mathbf{c}\mkern1mu{:}\mathbf{S}^\theta) \cdot d\vec{S}
        \label{force},
\end{align}
from which we verify that the PTh force field is composed of surface ($\vec{t}_\theta$) and body ($\vec{f}_\theta$) loads given by:
\begin{align}
        \vec f_\theta(\vec{r},t) &=  -\bm \nabla \cdot (\mathbf{c}\mkern1mu{:}\mathbf S^\theta)\label{vec_form2} ,\\
        \vec t_\theta(\vec{r},t) &= (\mathbf{c}\mkern1mu{:}\mathbf S^\theta)\cdot\hat{n}
        \label{vec_form1}.
\end{align}

The volume load $\vec{f}_\theta$ was used in past work on optomechanical PTh forces to provide an estimate of their magnitude~\cite{Schliesser2010CavityMicro-Resonators}. We demonstrate here that both surface and volume contributions are generally relevant in microscale devices, and must be considered for accurately describing dynamical backaction in optomechanical systems. 

In order to grasp the time-dependence of PTh forces, a constitutive relation  between the thermal strain and temperature field $\delta T(\vec{r},t)$ must be assumed, $\mathbf S^\theta = \bm \alpha \delta T(\vec{r},t)$, where $\bm \alpha$ is the thermal expansion tensor. The temporal analysis can be simplified by expanding $\delta T(\vec{r},t)$ in multiple thermal modes~\cite{Panuski2020FundamentalMicrocavities}, $\delta \Tilde{T}_k(\vec{r})$, with different relaxation constants, $\tau_k$, as: $\delta T(\vec{r},t) = \sum_k \theta_k(t) \delta \Tilde{T}_k(\vec{r})$, where $\theta_k(t)$ is the $k$-thermal mode amplitude. This procedure is described in detail in section S2 of the Supplemental Material. In this framework, the PTh force for the $n$-th-mechanical can be written as a sum of the contribution from multiple thermal modes as $F^{\theta}_n = \sum_k \Lambda^{\theta}_{k,n} \theta_k$, where $\Lambda^{\theta}_{k,n} = \int {\mathbf{S}}^x_n \, \mkern1mu{:} \, (\mathbf{c} \, \mkern1mu{:} \, \bm \alpha) \, \delta T_k \, dV $. Similarly, the surface and volume contributions can be decomposed in terms of $\Lambda^{\theta\text{-Vol.}}_{k,n}$ and $\Lambda^{\theta\text{-Sur.}}_{k,n}$.

Due to the optical drive (at frequency $\omega_l$), the evolution of the thermal amplitudes $\theta_k(t)$ is given by:
\begin{equation}
    \dot{\theta}_k = -\frac{1}{\tau_k}\theta_k + \frac{\hbar \omega_l \kappa_\text{abs}R^\theta_k}{\tau_k}|a(t)|^2,
\label{thermal_amplitude}
\end{equation}
where the thermal response to the optical heat source is modeled through the thermal relaxation time $\tau_k$, the thermal resistance $R^\theta_k$ and the optical absorption rate $\kappa_\text{abs}$. The electromagnetic mode amplitude $a(t)$ is normalized such that $|a(t)|^2$ is the number of circulating photons in the cavity. In this analysis, we neglect thermoelastic damping heat relative to the optical absorption one. Although equations similar to Eq.~\ref{thermal_amplitude} appeared in the literature~\cite{Guha2017HighSemiconductors, Woolf2013OptomechanicalMembranes}, their use was performed considering a single ``effective" thermal relaxation time obtained from a fit of experimental data regarding the magnitude of the PTh forces or from a frequency-resolved measurement of the resonator's thermo-optical response. Those two methods often yield contrasting results ~\cite{Guha2020ForceSelf-Oscillator}; our model indicates that this comes in place due to different thermal modes being relevant to the thermo-mechanical and thermo-optical processes. For instance, in our devices, the thermo-optical response is well represented by the fundamental thermal mode while this approximation for PTh effects yields approximately $200\%$ error in the overall force.  We stress that in the case of ``effective" thermal relaxation times no formal derivation of Eq.~\ref{thermal_amplitude} can be given, and its use is only justified from a phenomenological point of view.

\section*{Numerical modeling}
\label{sec:NM}

\textit{Static case --} In order to numerically validate the present model, we first consider the case of static thermal deformations on a GaAs on Al$_{0.7}$Ga$_{0.3}$As (\SI{250}/\SI{2000} {\nano\metre}) microdisk with radius $R =\SI{6}{\micro\metre}$ and pedestal radius $R_\text{ped} = \SI{0.75}{\micro\metre}$. The first radial order optical TE mode of the disk is used as a heat source that drives a stationary temperature field in the structure, as shown in Fig.~\ref{fig1} \textbf{b)}. Due to the static nature of this problem, thermal modal analysis is not necessary, such that the full thermal field is used in all calculations following.

We use finite element method (FEM) calculations to compare the thermal displacement $\vec{u}_\theta$ predicted by the derived PTh force field (PTh) to a fully-coupled thermoelastic model in COMSOL Multiphysics\copyright. The fully-coupled model calculations are carried out in the linear elastic approximation, in consistency with the hypothesis used in our derivation. For completeness, we further calculate the temperature-induced displacement resulting from the volume force (PTh - Vol.) alone, as shown in Fig.~\ref{fig1}~\textbf{c)}; the thermal displacement field components along $\hat r$ and $\hat z$ directions are shown in Fig.~\ref{fig1}\textbf{d)}. Our PTh force formulation, which includes both surface and volume contributions, accurately reproduces the fully-coupled model thermal displacement field, with major deformations present near the edge of the disk. This is in stark contrast with the volume-only PTh force calculations, where deformations are mostly confined to the pedestal region. This discrepancy indicates that thermo-mechanical coupling calculations can be critically affected by the existence of the surface PTh force in microphotonic structures.

\textit{Dynamic case --} To study how surface and volume loads may affect an optomechanical system, we solve the coupled-mode equations for optical, thermal and mechanical amplitudes in the frequency domain (S3 of the Supplemental Material), under a continuous-wave optical excitation which allows the linearization of these equations around an average amplitude. This procedure allows one to evaluate and compare the average photothermal force per photon obtained from surface and volume contributions. Neglecting optical resonance frequency shifts due to temperature variations, the lumped bolometric force (PTh) per photon is given by:
\begin{equation}
    F^{\theta}_n (|a|^2 = 1) = \hbar \omega_l \kappa_\text{abs} \sum_k \frac{R^\theta_k \Lambda^{\theta}_{k,n} \chi^{\theta}_k (\Omega_n)}{\tau_k},
    \label{eq:force_p_photon}
\end{equation}


\begin{figure}[H]
\centering
\includegraphics[width = 8.5cm]{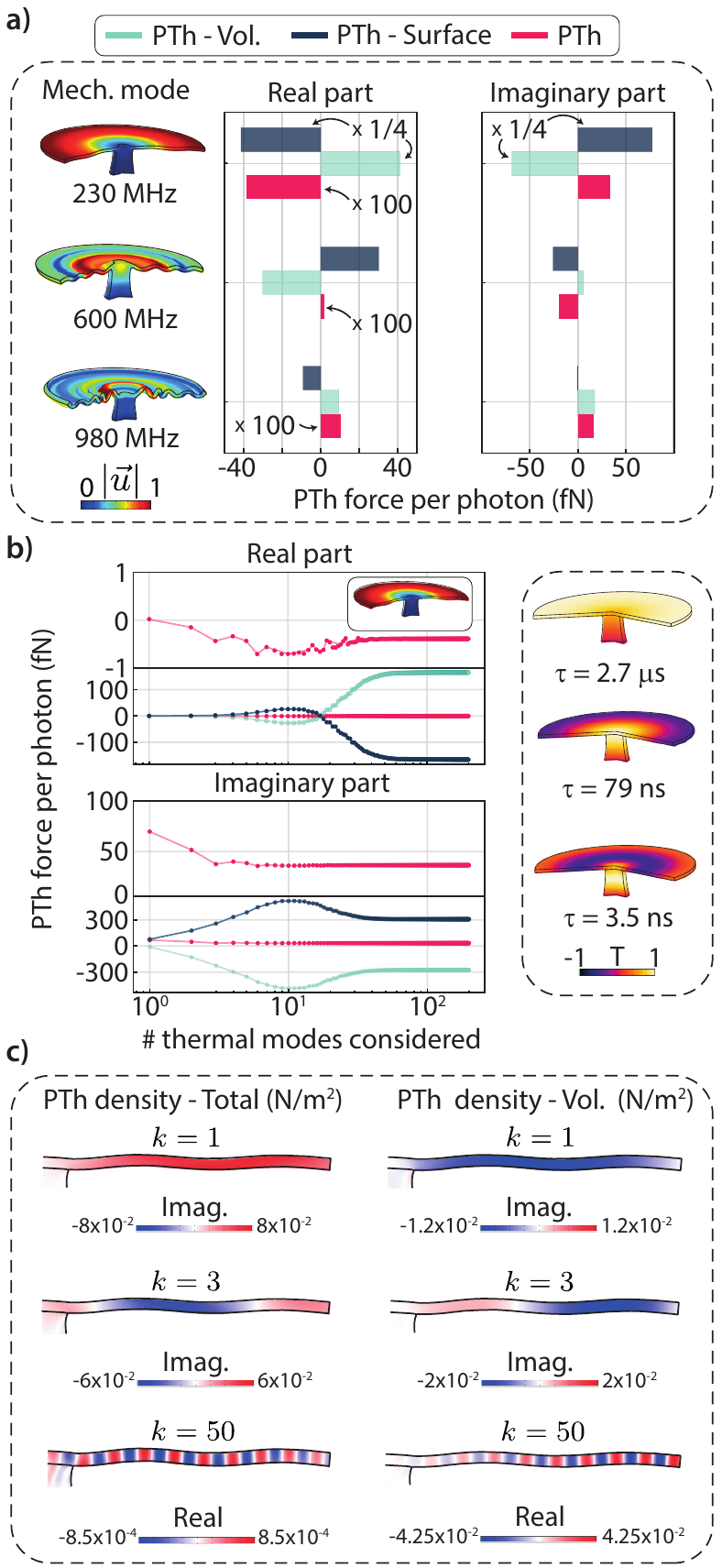}
\caption{\small{\textbf{a)} Average force per photon (real and imaginary parts) for three different mechanical modes due to the volume (PTh - Vol.), surface (PTh - Surface) and full (PTh) photothermal forces. The volume and surface forces terms acting on the $\SI{230}{\mega\hertz}$ mechanical breathing mode are re-scaled ($1/4$) for presentation purposes.  The phase acquired by the photons --  that depends on the laser-cavity detuning and optical linewidth -- is not considered here. \textbf{b)} Surface, volume and total PTh forces as functions of the number of thermal modes considered. Figures on top display the total PTh force in a suitable scale. Right corner: three lowest order thermal modes. \textbf{c)} Profiles of the total and volume component of the PTh force density within the microdisk for three different thermal modes $\delta T_k$, $k = 1,\,3,\,50$. These quantities are defined respectively as $2\pi r$ times the integrands of Eqs.~\ref{main} and \ref{force_vol}. The colorbars display the amplitude of the dominant quadrature (real/imaginary) of the force. The Imag./Real component ratios for the $k =1,\,3,\,50 $ modes considered are respectively given by $ 4\times 10^3,\,50,\,0.47$.}} 
\label{fig2}
\end{figure}

\noindent where $\chi^{\theta}_k (\Omega_n)$ is the thermal susceptibility of the $k$-th thermal mode evaluated at the angular frequency of the $n$-th mechanical mode.

The surface (PTh - Surface) and volume (PTh - Vol.) contributions are shown in Fig.~\ref{fig2} \textbf{a)} for three different mechanical modes. Here, we include the effects of fluctuations in the optical frequency due to the thermo-optical effect. Calculations involving the volume force alone not only would drastically overestimate the total PTh forces but would also carry a flipped sign with respect to the correct results. In our case, localized losses such as surface absorption do not modify the PTh response appreciably -- as shown in S4 of the Supplemental Material --, hence we assume that the absorptivity of GaAs is homogeneous and isotropic throughout the device. This result holds if the PTh response is dominated by low-order thermal modes that are essentially homogeneous in the region of confinement of the optical mode. If this is not the case, a thorough characterization of the nature of absorptive losses is necessary to obtain accurate predictions from the model. 

Surface and volume contributions are obtained by replacing $\Lambda^{\theta}_{k,n}$ with $\Lambda^{\theta-\text{Sur.}}_{k,n}$ and $\Lambda^{\theta-\text{Vol.}}_{k,n}$ in Eq.~\ref{eq:force_p_photon}. Since the $\chi^{\theta}_k (\Omega_n)$ are complex numbers, forces are composed of real and imaginary parts; the latter is largely dominant in the total PTh forces. Physically, this phenomenon is related to the relatively large thermal relaxation times $1/\tau_k\ll \Omega_n$ of the relevant thermal modes,  which cause their response to lag behind the mechanical oscillations. The optical absorption rate was chosen to be $\kappa_\mathrm{abs}/(2\pi) =\SI{1}{\GHz}$ following state-of-the-art experiments on GaAs microdisks~\cite{Parrain2015OriginResonators}. The total loss rate ($\kappa = \kappa_e + \kappa_\mathrm{abs}+\kappa_\mathrm{non-abs}$) is $\kappa/(2\pi)\approx \SI{1.93}{\giga\hertz}$, with extrinsic coupling rate (i.e. coupling to a waveguide) $\kappa_e/(2\pi)\approx \SI{0.48}{\giga\hertz}$. These numbers yield a loaded quality factor $Q_\text{opt} \approx 10^5$; all other parameters are obtained through FEM simulations, where first order  TE optical mode was considered.

The summation in Eq.~\ref{eq:force_p_photon} raises a question on the number of thermal modes that must be accounted for to correctly evaluate $F_n^\theta$. In Fig.~\ref{fig2} \textbf{b)} we consider the $\SI{230}{\mega\hertz}$ mechanical breathing mode, and calculate the contributions of the surface and volume components to the total PTh force per photon as a function of the number of thermal modes considered, ordered by decreasing $\tau_k$ . The total PTh response -- largely dominated by the imaginary component -- takes only $6$ thermal modes to converge reasonably, whereas the volume and surface forces take $\approx 40$ thermal modes.  This feature arises from the fact that in high order ($k>6$) thermal modes volume and surface terms yield opposite contributions that approximately cancel each other. Physically, this comes in place because the temperature profiles of high order thermal modes are associated with rapid spatial oscillations within the microdisk; as the spatial frequency increases, strong temperature-gradient forces (Eq.~\ref{force_vol}) are generated. This behavior, however, is not verified for the total force in Eq.~\ref{main}, in which the gradient operation appears acting on the elastic deformations through the quantity $\mathbf{S}^x$. Consequently, as the order of the thermal modes is increased, the integrand in Eq.~\ref{force_vol} overcomes its counterpart in Eq.~\ref{main}, indicating that volume forces should become larger than the total force (in absolute terms). The only way such phenomenon can be observed is if surface and volume contributions counterbalance each other. We illustrate this in Fig.~\ref{fig2}\textbf{c)}, where the profiles of the total and volume pressures per photon are displayed for three different thermal modes $\delta T_k$, $k = 1,\,3,\,50$. Remarkably, in the $k=1, \,3$ modes, the total and volume photothermal pressures are of similar amplitude, whereas for $k = 50$, the volume component is almost two orders of magnitude larger than its counterpart. Lastly,  since volume and surface terms yield opposite contributions even for the dominant low order thermal modes, surface engineering may emerge as a route for the enhancement or even cancellation of PTh forces.


\begin{figure}[ht!]
\includegraphics[width = 8.5cm]{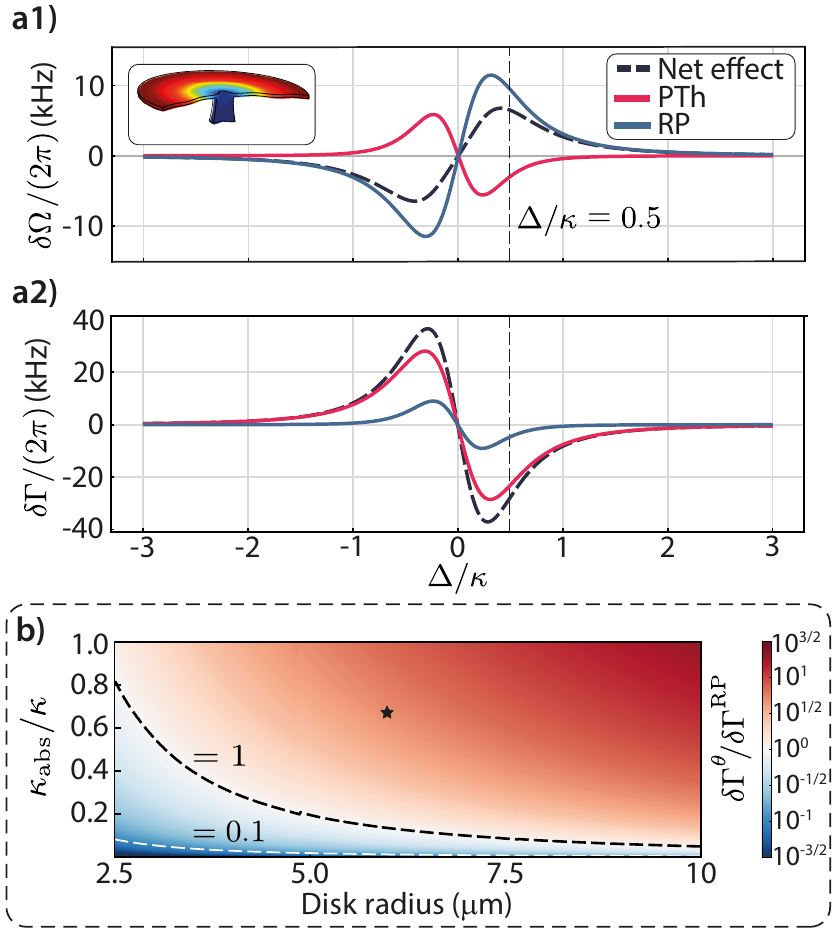}
\caption{\small{  Radiation pressure (RP) and photothermal (PTh) backaction-induced mechanical \textbf{a1)} frequency and \textbf{a2)} linewidth shifts for the mechanical breathing mode in Fig.~\ref{fig2}. \textbf{b)} Ratio between mechanical damping rate modifications $\delta\Gamma^\theta/\delta\Gamma^\text{RP}$ induced by PTh and RP forces evaluated at half optical linewidth ($\Delta = \kappa/2)$. The star marker depicts the device in \textbf{a1)} and \textbf{a2)}.}}
\label{fig3}
\end{figure}

We now turn our attention to the complete optomechanical interaction considered in this work, composed of radiation pressure and photothermal forces. Both contribute independently to an effective optomechanical backaction and must be considered for a correct description of the effects that will be studied in our experiment. We consider the same device as in Fig.~\ref{fig2}, with a mechanical breathing mode at \SI{230}{\MHz} and \SI{50}{\micro\watt} incident power. In Figs.~\ref{fig3} \textbf{a1)} and \textbf{a2)} the PTh and RP backaction curves are displayed. For the RP calculations, both photoelastic~\cite{Balram2014MovingResonators} and moving boundary contributions~\cite{Johnson2002PerturbationBoundaries} are considered. While RP dominates the optically-induced frequency shift, cooling and amplification are largely dominated by PTh forces. This is due to slow thermal responses (when compared to the mechanical periods) yielding PTh forces out-of-phase with respect to the mechanical oscillations, which favors mechanical linewidth modification processes. This is a key feature that is explored in our experiments. Importantly, for GaAs microdisks, PTh and RP effects add constructively in cooling/amplification processes. In Fig.~\ref{fig3} \textbf{b)} the ratio of PTh and RP cooling at $\Delta = 0.5 \kappa$ is evaluated as a function of $\kappa_\text{abs}$ and the disk radius; for these calculations, the pedestal radius $R_\text{ped} = \SI{0.75}{\micro\metre}$ is kept fixed. Such diagram can be used as a tool for choosing geometries in order to maximize or suppress PTh effects: while larger disks display PTh-dominated dynamical backaction (red region), in smaller disks -- where optical and mechanical modes are more tightly confined and with larger overlap -- RP interaction prevails (blue region). The marker displays the parameters used in Figs.~\ref{fig3} \textbf{a1)}, \textbf{a2)}. 

\begin{figure*}[ht!]
\centering
\includegraphics[width = 16.0cm]{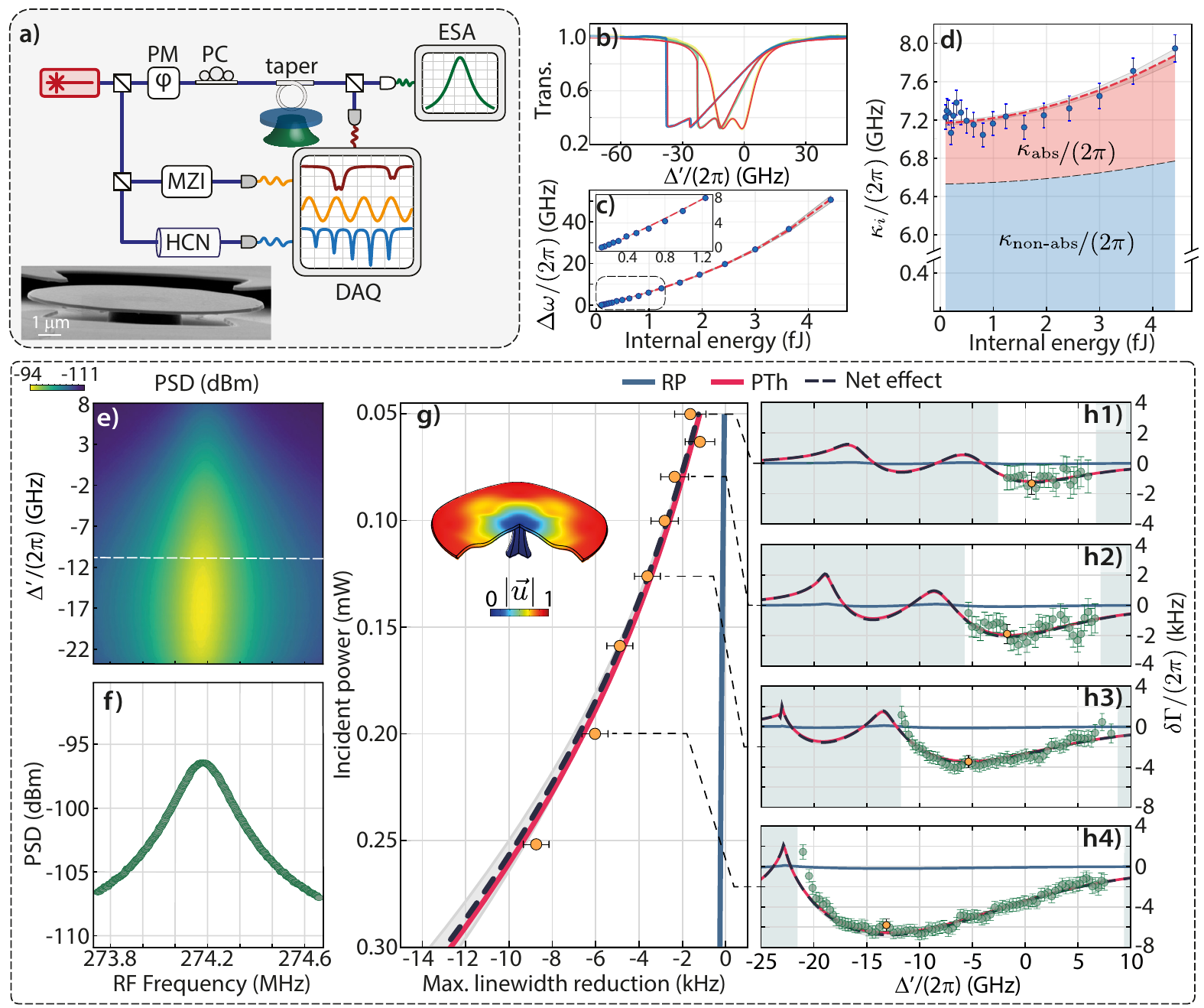}
\caption{\small{\textbf{a)} Experimental setup for optomechanical characterization.  \textbf{b)} Transmission spectra for different incident powers: $\SI{15.9}{\micro\watt}$ (yellow), $\SI{126}{\micro\watt}$ (green) and $\SI{252}{\micro\watt}$ (blue); along with theoretical curves (red) calculated using the nonlinear response characterization. \textbf{c)} Nonlinear optical dispersion. Inset: Low power dispersion, with approximately linear scaling as a function of internal energy. \textbf{d)} Total intrinsic loss as function of internal energy. \textbf{e)} Mechanical spectrogram for an incident power of $\SI{250}{\micro\watt}$ as a function of detuning ($\Delta$). \textbf{f)} Mechanical power spectral density evaluated at the white dashed line in \textbf{e)}. \textbf{g)} Measured and calculated contributions to the maximal $\delta\Gamma$ as a function of several input powers. \textbf{h1)-h4)} Comparison between measured and calculated $\delta\Gamma$ as a function of cold-cavity detuning. The shape of these curves is related to the optical doublet under analysis. Points in \textbf{g)} are obtained from the results in \textbf{h1)-h4)}.}}
\label{fig4}
\end{figure*}

\section*{Experimental results} The effectiveness of the thermodynamic description is tested by monitoring the modification on the mechanical linewidth of a cavity optomechanical system consisting of a $R \approx (\text{5.1} \pm \text{0.1})\SI{}{\micro\metre}$, $R_\text{ped}\approx (\text{0.6} \pm \text{0.1})\SI{}{\micro\metre}$  GaAs/Al$_{0.7}$Ga$_{0.3}$As (\SI{250}/\SI{2000} {\nano\metre}) microdisk. The experimental setup used to characterize both the mechanical and optical spectra is shown in Fig.~\ref{fig4} \textbf{a)}. A scanning electron microscope (SEM) image of the fabricated device is also presented. Light emitted by a tunable laser source is coupled in and out the resonator through a tapered fiber loop. The output from the cavity is collected at both fast and slow photodetectors. The fast response is fed into an electrical spectrum analyzer (ESA), while the slow signal is collected by an analog-to-digital converter (DAQ). A Mach-Zehnder interferometer (MZI) and hydrogen cyanide reference gas cell (HCN) are used for the calibration of the cavity's optical response. A detailed description of the experimental setup is found in S5 of the Supplemental Material.

A thorough optical characterization of the device is necessary in order to calibrate both nonlinear losses and thermal frequency shift, both crucial to accurately predict backaction effects at high incident powers. We monitor the optical transmission spectrum of the cavity for various incident powers, as illustrated in Fig.~\ref{fig4} \textbf{b)}. The cold-cavity transmission yields intrinsic and extrinsic optical damping rates of $\kappa_i/(2\pi) = \SI{7.0}{\GHz}$ and $\kappa_e/(2\pi) = \SI{4.2}{\GHz}$. Assuming that the coupling to the fiber taper remains constant during the measurements, power-dependent changes in the transmission can be traced back to recover the nonlinear losses and internal optical energy of the resonator ($U$)~\cite{Barclay2005NonlinearTaper,Borselli2007AccurateDevices}. Nonlinear frequency shift is directly obtained through joint calibration with the MZI and tracking of the resonance shift ($\Delta\omega$). Nonlinear losses and frequency shift data are then simultaneously adjusted to polynomial curves to obtain absorptive and non-absorptive optical dissipation rates as a function of the energy in the resonator. The polynomial approximation is valid for sufficiently low input powers, where only terms up to $O(U^3)$ in internal energy are enough to describe our results.  A comprehensive guide for this analysis is found in S6 of the Supplemental Material. In Fig.~\ref{fig4} \textbf{c)} we show $\Delta\omega$ as a function of the internal energy in the resonator. The inset shows the dispersion for low incident powers, critical for determining the portion of the cold-cavity losses with absorptive nature. Fig.~\ref{fig4} \textbf{d)} displays the total optical dissipation rate of the system split in absorptive ($\kappa_\text{abs}$) and non-absorptive ($\kappa_\text{non-abs}$) parts.

We measure the back-action effects by monitoring the mechanical mode spectrum through the RF power spectrum, which is recorded for a range of positive (blue) laser-cavity detuning, resulting in spectrograms similar to Fig.~\ref{fig4} \textbf{e)}. From a Lorentzian fit (shown in Fig.~\ref{fig4}~\textbf{f)}) both mechanical frequency ($\delta\Omega$) and linewidth ($\delta\Gamma$) changes are obtained and the latter is compared with the predictions of the PTh and RP models previously discussed. For the tested device, RP yields negligible contribution, demonstrating the role of distinct backaction mechanisms in explaining the observed phenomena. The optical mode excited in our measurements is identified through its free-spectral range (FSR), consistent with the $6$-th order TE optical mode (S7A of the Supplemental Material). 

The estimated photothermal response of the system is obtained through a combination of FEM simulations for a mechanically anisotropic GaAs microdisk (S7B of the Supplemental Material) and the experimental nonlinear loss and dispersion described above. Importantly, our FEM results -- where the parameters $R_k^\theta$ and $\Lambda_k^\theta$ were evaluated -- require only thermal and thermo-elastic material properties as input, all of which are well established in the literature~\cite{Adachi1985}. Fig.~\ref{fig4} \textbf{g)} exhibits the comparison between the measured maximal $\delta\Gamma$ and its theoretical estimate as a function of the incident powers on the cavity. Those values are obtained from measurements of $\delta\Gamma$ as a function of the laser to cold-cavity detuning, $\Delta'$, (i.e. $\Delta' = 0$ refers to the cold-cavity resonance frequency), exemplified in Figs.~\ref{fig4}~\textbf{h1)-h4)} for four different input powers. Details on the collection and analysis of data regarding the mechanical response are found in S8 of the Supplemental Material, along with the treatment of the stiffening of the mechanical oscillator ($\delta\Omega$),  which is dominated by a static temperature softening of GaAs~\cite{Gil-Santos2013OpticalEffects}, red-shifting the mechanical frequency up to $-\SI{20}{\kilo\hertz}$. As expected, the uncertainty in $\delta\Gamma$ becomes more relevant at low input powers, since in that case the optomechanical transduction is less efficient.

Excellent agreement between our prediction (curves) and experiment (markers) is found over the whole range of detuning measured in Figs.~\ref{fig4}~\textbf{h1)-h4)}. Despite the encouraging nature of our results, we stress that, as discussed in S4 of the supplementary material, in the microdisk geometry it is hard to distinguish bulk from surface absorption. Experiments with more sophisticated resonators, in which PTh effects are sensitive to the spatial distribution of the absorption, provide an interesting route for testing further our predictions.


\section*{Conclusion}  In summary, we have proposed and verified experimentally a  model for the photothermal forces acting on cavity optomechanical systems derived through thermodynamic considerations. The theoretical estimates were shown to display remarkable agreement with our measurements, thus providing a solid route to design the thermo-optomechanical response in nanomechanical resonators. Also, the modal treatment for the thermal response is a significant step towards thermal engineering in the broad field of nanophotonics, and paves the way for a new class of experiments where those effects are tailored to interest. Finally, although GaAs based devices were taken as an example, we emphasize photothermal forces can be appreciable in other platforms and geometries and that the content of can provide insight in those cases. 

\section*{ACKNOWLEDGMENTS}
The  authors  would  like  to  acknowledge  CCSNano-UNICAMP  for  providing  the  micro-fabrication infrastructure and CMC Microsystems for providing access to MBE epitaxy and the GaAs wafers. This work was supported by São Paulo Research Foundation (FAPESP) through grants 2019/09738-9, 2020/06348-2, 2017/14920-5, 2016/18308-0, 2017/19770-1, 2018/15580-6, 2018/15577-5, 2018/25339-4, Coordena{\c c}\~ao de Aperfei{\c c}oamento de Pessoal de N{\'i}vel Superior - Brasil (CAPES) (Finance Code 001), Conselho Nacional de Desenvolvimento Científico e Tecnológico through grants 425338/2018-5, 310224/2018-7, 465469/2014-0, Financiadora de Estudos e Projetos (Finep) and the Natural Sciences and Engineering Research Council (NSERC) of Canada. 

\section*{DATA AVAILABILITY}
FEM and scripts files for generating each figure are available at Ref.~\cite{zenodo_bolometric}. Additional data that support the findings of this study are available from the corresponding author upon reasonable request.


%

\end{document}


\title{Supplemental Material: Accurate modeling and characterization of photothermal forces in optomechanics}

\author{Andr\'{e} G. Primo$^{1,2,*}$}
\author{Cau\^{e} M. Kersul$^{1,2,*}$}
\author{Rodrigo Benevides$^{1,2}$}
\author{Nat\'{a}lia C. Carvalho$^{1,2}$}
\author{Micha\"{e}l M\'{e}nard$^{3}$}
\author{Newton C. Frateschi$^{1,2}$}
\author{Pierre-Louis de Assis$^{1}$}
\author{Gustavo S. Wiederhecker$^{1,2}$}
\author{Thiago P. Mayer Alegre$^{1,2,\dagger}$}

\affiliation{\text{1 - Applied Physics Department, Gleb Wataghin Physics Institute, University of Campinas, Campinas, SP, Brazil}\\
\text{2 - Photonics Research Center, University of Campinas, Campinas, SP, Brazil}\\
\text{3 - Department of Computer Science, Universit\'{e} du Qu\'{e}bec \`{a} Montr\'{e}al, Montr\'{e}al, Canada}\\
\text{* These authors contributed equally for this work.}\\
\text{$\dagger$ Corresponding author: alegre@unicamp.br}}

\date{\today}

\maketitle
\tableofcontents

\newpage
\section*{Table Summary}

Tab.\ref{tab:parameters} summarizes all experimental and simulated parameters used throughout the main and supplemental text. Each parameter used on our model is also presented with its value, source (the way they were obtained), and the section where it is discussed.

\begin{center}
\LTcapwidth=\textwidth
\begin{table}[ht!]
\begin{tabular}{ |p{5cm}|c|c|p{3cm}|c| }
\hline
 Parameter description & Symbol & Value & Source & Section \\
 \hhline{|=|=|=|=|=|}
 First resonance's optical frequency & $\omega_c/(2\pi)$ & \SI{191.5418\pm0.0002}{THz} & Experiment & \ref{subsec:fitting_2}\\
 \hline
 Initial extrinsic loss & $\kappa_{e}/(2\pi)$ & \SI{4.197\pm0.002}{GHz} & Experiment & \ref{subsec:fitting_2} \\
 \hline
 Linear intrinsic loss & $\kappa_{i}/(2\pi)$ & \SI{7.33\pm0.01}{GHz} & Experiment & \ref{subsec:fitting_2} \\
 \hline
 Doublet splitting & $\delta\omega/(2\pi)$ & \SI{11.314\pm0.007}{GHz} & Experiment & \ref{subsec:fitting_2}\\
 \hline
 Correction to the linear losses & $\kappa_{\text{nl,0}}/(2\pi)$ & \SI{-0.16\pm0.02}{GHz} & Experiment & \ref{subsec:fitting_3} \\
 \hline
 First order nonlinear loss coefficient (TPA) & $\kappa'_1/(2\pi)$  & \SI{20}{MHz/fJ} & FEM simulations$^a$ & \ref{subsec:fitting_3} \\
 \hline
 Second order nonlinear loss coefficient & $\kappa'_2/(2\pi)$  & \SI{31\pm3}{MHz/fJ^2} & Experiment & \ref{subsec:fitting_3} \\
 \hline
 Correction to the linear dispersion & $\delta\omega_{nl,0}/(2\pi)$  & \SI{0.93\pm0.06}{GHz} & Experiment & \ref{subsec:fitting_3} \\
 \hline
 Thermo-optic shift per heating power (static source)$^b$ & $G^{\theta}_s R^{\theta}_s/(2\pi)$  & \SI{1695}{GHz/mW} & FEM simulations$^a$ & \ref{subsec:fitting_3} \\
 \hline
 Absorptive nonlinear losses ratio & $\eta_{nl}$ & \SI{0.66\pm0.01}{} & Experiment & \ref{subsec:fitting_3} \\
 \hline
 Cold cavity mechanical frequency & $\Omega$ & \SI{274.184\pm0.001}{MHz} & Experiment and FEM simulations$^a$ & \ref{subsec:optical_id_2} \\
 \hline
 Cold cavity mechanical loss & $\Gamma$ & \SI{175.0\pm0.2}{kHz} & PSD & \ref{subsec:om_response_2}\\
 \hline
 Optomechanical coupling & $G_x \sqrt{\frac{\hbar}{2m_{\text{eff}}\Omega}}$ & \SI{29.2\pm0.5}{kHz} & Experiment \cite{Gorodetsky}, and FEM simulations$^a$ & \ref{subsec:optical_id_2} \\
 \hline
 Response time of the different thermal modes & $\tau_k$ & - & FEM simulations$^a$ & \ref{subsec:freq_domain_1}\\
 \hline
 Thermal resistances & $R_{\text{k,j}}^{\theta}$ & - & FEM simulations$^a$ & \ref{subsec:freq_domain_1} \\
 \hline
 Thermo-elastic couplings & $\Lambda_{\text{k}}^{\theta}$ & - & FEM simulations$^a$ & \ref{subsec:freq_domain_1}\\
 \hline
 Thermo-optic couplings & $G_{\text{k}}^{\theta}$ & - & FEM simulations$^a$ & \ref{subsec:freq_domain_1} \\
 \hhline{|=|=|=|=|=|}
   \multicolumn{5}{|p{15cm}|}{$^a$ In the FEM simulations the material properties were taken from references, \cite{doi:10.1063/1.3533775,Adachi1985,DellaCorte2000Temperaturem,Balram2014MovingResonators,Soma1982ThermalInP}} \\
  \multicolumn{5}{|p{15cm}|}{$^b$ The static thermal response of our system is largely dominated by the fundamental thermal mode.} \\
 \hline
\end{tabular}
 \caption{All parameters used in our model, indicating both its value and how they were obtained. The value of parameters dependent on the different thermal modes are not shown here.}
\label{tab:parameters}
\end{table}
\end{center}


\section{Derivation of the photothermal coupling}\label{sec:derivation}

We discuss two different approaches to derive the photothermal coupling; the first is built from thermodynamic arguments and the second directly from the mechanical equation of motion. Our analysis assumes that all transport processes are quasi-static, in the sense that local equilibrium is assumed at each step. This hypothesis allows us to safely use definitions of quantities such as temperature and is valid within the mechanical frequencies considered in this work. Local heat transport is ultimately related to thermal phonons relaxation times which range around $10^{-11}\SI{}{s}$, much faster than the mechanical oscillations with periods around $10^{-9} \SI{}{s}$.

\subsection{Thermodynamic derivation}\label{subsec:derivation_1}

The total strain tensor ($\bm S$) in a thermoelastic body is given by:

\begin{equation}
    \bm S = \bm S^x+ \bm S^\theta = \hat{\bm \nabla} \vec{U},
    \label{straint}
\end{equation}
where $\bm S^\theta$ and $\bm S^x$ are the thermal and elastic components of the strain, respectively. The symbol $\hat{\bm \nabla} = \frac{1}{2}(\nabla+\nabla^T)$ denotes the symmetric gradient tensor operation and $\vec{U}$ is the displacement field. The thermal strain tensor is defined as:

\begin{equation}
    \bm S^\theta = \bm \alpha \delta T,
\end{equation}
where $\bm \alpha$ is the rank 2 thermal expansion tensor and $\delta T$ is the temperature field of the body. In an isotropic medium $\bm \alpha$ can be simplified to, $\alpha \mathcal{I}$, where $\alpha$ is the thermal expansion coefficient and $\mathcal{I}$ is the $3\times 3$ identity matrix.

Denoting $\bm c$ as the stiffness tensor, the stress tensor is obtained through the following constitutive relation:

\begin{equation}
\boldsymbol \sigma = \bm c:\bm S^x =\bm c:\bm S - \bm c:\bm S^\theta.
\label{stress_const}
\end{equation}
We argue that the additional stresses related to photothermal forces are necessarily related to the term $-\,\bm c:\bm S^\theta$. Evaluating the work per unit volume $\delta w$ done during an arbitrary deformation $\delta \vec{U}$ we get:
\begin{equation}
    \delta w = \frac{\partial \sigma_{ij}}{\partial X_j} \delta U_i,
\end{equation}
where $X_j$ ($j = 1,2,3$) denotes each of the cartesian coordinates and summation over repeated indexes is used. Considering the work in the entire volume where the deformation takes place we have:

\begin{equation}
    \delta W = \int \delta w \, dV = \int \frac{\partial \sigma_{ij}}{\partial X_j} \delta U_i \, dV.
\end{equation}
Using Gauss's theorem and integration by parts:

\begin{equation}
    \int \frac{\partial \sigma_{ij}}{\partial X_j} \delta U_i dV = -\int \sigma_{ij}\delta \frac{\partial U_i}{\partial X_j}dV + \int \sigma_{ij} \delta U_i dA_j.
\end{equation}
For a free surface ($ \boldsymbol \sigma \cdot  \hat{n} = 0$), the second integral on the RHS vanishes and we get:
\begin{equation}
    \delta w = - \sigma_{ij}\delta \frac{\partial U_i}{\partial X_j} = - \sigma_{ij}\delta S_{ij}.
    \label{work}
\end{equation}
where we have identified $\frac{\partial U_i}{\partial X_j}$ with the total strain. 

We are only interested in forces that effectively transfer energy from the thermal to the mechanical domain over a stationary cycle \footnote{Here, we define the stationary cycle period as the time necessary for the strain to return to its initial state.}, specifically, in the form of work. Substituting Eq.~\ref{stress_const} in Eq.~\ref{work} we get:

\begin{equation}
    \delta w= - c_{ijkl}S_{kl} \delta S_{ij}+ c_{ijkl}S^\theta_{kl} \delta S_{ij}.
\end{equation}

Integrating this expression over a mechanical cycle and noting that the first term  on the right-hand side is an exact differential -- and therefore vanishes --, we conclude that $c_{ijkl}S^\theta_{kl} dS_{ij}$ is responsible for feeding energy into the mechanical domain. Aiming at finding a linear theory, underpinned by modal analysis, we keep only terms to first order in temperature. The final expression for the work per unit volume, done by photothermal forces, is given by:

\begin{equation}
    \delta w^\theta = c_{ijkl}S^\theta_{kl} \delta S^x_{ij}.
\end{equation}
Integrating over the entire volume we obtain the total photothermal work:

\begin{equation}
    \delta W^\theta = \int{ (\mathbf{c}\mkern1mu{:}\mathbf{S}^\theta) \mkern1mu{:} \, \delta \mathbf{S}^x dV}.
\end{equation}

The elastic strain  can be decomposed in the mechanical modes $\vec{u}_n(\vec{r})$ of the system as $\mathbf{S}^x =   \sum_n x_n(t) \bm S^x_n$, where $x_n(t)$ is the $n$-{th} mode amplitude (i.e. the $\vec{u}_n(\vec{r})$ are normalized such that its global maximum is set to be $1$) and $\bm S^x_n = \hat{\bm \nabla} u_n(\vec{r})$ . Using the modal amplitudes $x_n$ as generalized coordinates in the sense of analytical mechanics, the generalized force on the $n$-th acoustic mode is:

\begin{equation}
   \frac{\partial W^\theta}{\partial x_n} = F^{\theta}_{n}(t) = \int{ \frac{\partial \mathbf{S}^x}{\partial x_n}\mkern1mu{:}(\mathbf{c}\mkern1mu{:}\mathbf{S}^\theta) dV.}
\end{equation}
Or simplifying:

\begin{equation}
   F^{\theta}_{n}(t) = \int{ \mathbf S^x_n\mkern1mu{:}(\mathbf{c}\mkern1mu{:}\mathbf S^\theta) \, dV},
\end{equation}
and explicitly in terms of mechanical modes and the temperature field:

\begin{equation}
   F^{\theta}_{n}(t) = \int{ \hat{\bm \nabla} u_n(\vec{r})\mkern1mu{:}(\mathbf{c}\mkern1mu{:}\bm \alpha) \, \delta T (\vec{r},t) dV}.
    \label{eq:force_final_n}
\end{equation}

\subsection{Derivation from the mechanical equation of motion}\label{subsec:derivation_2}
We start with the mechanical equation of motion:

\begin{equation}
    \rho \ddot{\vec{U}}=\bm \nabla\cdot \boldsymbol \sigma,
    \label{stress_gen} 
\end{equation}
with the boundary condition:

\begin{equation}
    \boldsymbol \sigma \cdot \hat{n} = 0.
    \label{bc1}
\end{equation}

In view of Eq.~\ref{stress_const}, these equations can be recast as:

\begin{equation}
    \rho \ddot{\vec{U}} = \bm \nabla\cdot\big( \bm c \mkern1mu{:} \bm S \big) - \bm \nabla\cdot\big( \bm c \mkern1mu{:} \bm S^{\theta} \big),
    \label{stress_gen_2} 
\end{equation}
with the boundary condition:

\begin{equation}
    \big( \bm c \mkern1mu{:} \bm S \big) \cdot \hat{n} = \big( \bm c \mkern1mu{:} \bm S^{\theta} \big) \cdot \hat{n}.
    \label{bc2}
\end{equation}
The thermal strain in the above equations can be pictured to induce both an external volume load $\vec f_\theta = - \bm \nabla\cdot\big( \bm c \mkern1mu{:} \bm S^{\theta} \big)$, and a surface traction $\vec t_\theta = \big( \bm c \mkern1mu{:} \bm S^{\theta} \big) \cdot \hat{n}$. 

In order to perform the modal decomposition of the photothermal force it is important to consider the mechanical eigenmodes of our device. These are solutions of:

\begin{equation}
    -\rho\Omega_{n}^2\vec{u}_n = \bm \nabla\cdot\big( \bm c \mkern1mu{:} \bm S^x_n \big),
    \label{eigenvalue_set}
\end{equation}
with free (untractioned) boundary conditions:

\begin{equation}
\bm S^x_n \cdot \hat n = 0,
\label{bc3}
\end{equation}
where the strain is given by $\bm{S}^x_n = \hat{\bm \nabla} \vec{u}_n$, and $\Omega_{n}$ is the angular mechanical frequency of the $n$-th mechanical mode.

Projecting Eq.~\ref{stress_gen_2} in a mechanical mode $\vec{u}_n$ and performing a volume integration, we get:

\begin{equation}
    \int \rho \vec{u}_n \cdot \ddot{\vec{U}} \, dV=\int \vec{u}_n \cdot \Big( \bm \nabla\cdot \big( \bm c\mkern1mu{:}\bm S\big) \Big)  \, dV - \int \vec{u}_n \cdot \Big( \bm \nabla\cdot\big( \bm c \mkern1mu{:} \bm S^{\theta} \big) \Big) \, dV .
    \label{projection}
\end{equation}

Integration by parts of the first term on the right-hand side yields:

\begin{equation}
\int \vec{u}_n \cdot \Big( \bm \nabla\cdot\big(\bm c\mkern1mu{:} \bm S \big)\Big)  \, dV = \int \vec{u}_n \cdot \Big(\bm c\mkern1mu{:} \bm S\Big) \cdot \hat{n}\, dS - \int \vec{U} \cdot \Big(\bm c\mkern1mu{:} \bm S^x_n \Big) \cdot \hat{n}\, dS +\int \vec{U} \cdot \Big( \bm \nabla\cdot \big(\bm c\mkern1mu{:} \bm S^x_n \big)\Big) \, dV .
\label{term_decomposition}
\end{equation}

Using Eqs.~\ref{bc2} and \ref{bc3} to rewrite the surface integrals and also using Eq.~\ref{eigenvalue_set} to recast the volume integral on the right-hand side,  Eq.~\ref{term_decomposition} reads:

\begin{equation}
\int \vec{u}_n \cdot \Big(\bm \nabla\cdot(\bm c\mkern1mu{:} \bm S)\Big)  \, dV = \int \vec{u}_n \cdot \big( \bm c \mkern1mu{:} \bm S^{\theta} \big) \cdot \hat{n} \, dS - \Omega_{n}^2 \int \rho \vec{U} \cdot \vec{u}_n \, dV .
\label{term_decompostion_final}
\end{equation}

Substituting Eq.~\ref{term_decompostion_final} back in Eq.~\ref{projection}:

\begin{equation}
    \int \rho \vec{u}_n \cdot \ddot{\vec{U}} \, dV = - \Omega_{n}^2 \int \rho \vec{U} \cdot \vec{u}_n \, dV + \int \vec{u}_n \cdot \big( \bm c \mkern1mu{:} \bm S^{\theta} \big) \cdot \hat{n} \, dS - \int \vec{u}_n \cdot \Big( \bm \nabla\cdot\big( \bm c \mkern1mu{:} \bm S^{\theta} \big) \Big) \, dV .
\end{equation}

Finally, since the first term in the RHS is only defined inside the device volume, we can then perform a modal expansion of the displacement field as $\vec{U} = \sum_n x_n(t)\vec{u}_n(\vec{r})$. Using the orthogonality relations for the mechanical modes, $\int \rho \vec{u}_n \cdot \vec{u}_m\, dV = m_{\text{eff},n} \delta_{n,m}$, where $m_\text{eff}$ is effective mass, we get:

\begin{equation}
    m_\text{eff,n} \big( \ddot{x}_n + \Omega_{n}^2 \, x_n \big)= \int \vec{u}_n \cdot \big(\bm c \mkern1mu{:} \bm S^{\theta} \big) \cdot \hat{n} \, dS - \int \vec{u}_n \cdot \Big( \bm \nabla\cdot\big( \bm c \mkern1mu{:} \bm S^{\theta} \big) \Big) \, dV ,
    \label{projection_2}
\end{equation}
Usually, a damping term $\Gamma_{n} \dot x_n$ is heuristically introduced on the left hand side, but will be omitted for the moment.

The surface and the volume contributions to the photothermal force can be assembled in a single volume integral:
 \begin{equation}
    m_\text{eff,n}\big(\ddot{x}_n+\Omega_{n}^2 x_n\big)= \int \bm S^x_n \mkern1mu{:}\, \big( \bm c \mkern1mu{:} \bm S^\theta \big) \, dV
\end{equation}
Which agrees with the thermodynamic treatment previously discussed. In previous works the surface load, which is evident in our analysis, was ignored.

\section{Thermal mode analysis}\label{sec:thermal_modes}

The heat equation for diffusive heat transport reads:

\begin{equation}
    c_p \rho \partial_t \delta T(\vec{r},t) =  \nabla \cdot \big(k_{\text{th}} \nabla \delta T(\vec{r},t)\big) + \dot{Q}(\vec{r},t),
    \label{heat}
\end{equation}
where $c_p$, $\rho$ and $k_{\text{th}}$ are respectively the specific heat, mass density and thermal conductivity of the medium. $\dot Q$ is the heat generation in the structure, which will be associated with optical absorption.

In the absence of heat sources, this equation is separable and admits solutions of the form $\delta T(\vec{r},t) =  \delta \Tilde T(\vec{r})e^{-t/\tau}$ leading to a generalized eigenvalue equation for the thermal modes:

\begin{equation}
    \nabla \cdot \big(k_{\text{th}} \nabla \delta \Tilde T(\vec{r})\big) =  -\frac{1}{\tau}c_p\rho \delta \Tilde T(\vec{r}).
\end{equation}

Since the operators $\nabla \cdot \big(k_{\text{th}} \nabla)$ and $c_p\rho$  are both Hermitian and the latter is positive definite, the equation above is associated with a complete set of eigenmodes, $\delta T(\vec{r},t) = \sum_k \theta_k(t) \delta \Tilde{T}_k(\vec{r})$. Applying  this expansion to Eq.~\ref{heat} yields:

\begin{equation}
    c_p\rho  \sum_k\dot{\theta}_k(t)\delta \Tilde T_k(\vec{r}) = \sum_k [\theta_k \nabla \cdot \big ( k_{\text{th}} \nabla \delta \Tilde T_k(\vec{r}) \big)] + \dot{Q}(\vec{r},t).
\end{equation}

Assuming fixed temperature or insulating boundary conditions, one derives orthogonality relations for the thermal eigenmodes:

\begin{equation}
    \int{ c_p\rho \delta \Tilde T_k(\vec{r}) \delta \Tilde T_l(\vec{r})dV = C_{k}\delta_{k,l}}, \quad \text{with} \,\, C_k \neq 0,
\end{equation}
which allows the time evolution of any thermal mode to be written as:

\begin{equation}
    \dot{\theta}_k = -\frac{1}{\tau_k}\theta_k + \frac{\int{\dot{Q}(\vec{r},t)\delta \Tilde T_k(\vec{r})dV}}{\int{c_p\rho (\delta \Tilde T_k)^2 dV }}.
    \label{theta-evo-int}
\end{equation}

Applying the thermal mode expansion over the expression for the lumped photothermal force, Eq.~\ref{eq:force_final_n}, allows us to write $F^{\theta}_{n} (t) = \sum_k F^{\theta}_{k,n} (t)$, where:

\begin{equation}
       F^{\theta}_{k,n}(t) = \big[\int{ \hat {\bm\nabla} \vec{u}_n\mkern1mu{:}\big(\mathbf{c}\mkern1mu{:}\bm \alpha \big) \delta \tilde T_k(\vec{r}) dV} \big] \theta_k(t) = \Lambda_{k,n}^{\theta} \theta_k(t),
       \label{force_final_form}
\end{equation}
where the coefficient $\Lambda_{k,n}^\theta$ has units of N/K.

We are interested in heating due to optical absorption. Assuming that only one optical mode is excited and denoting the local absorptive (optical) loss rate  as $\kappa_{\text{loc}}(\vec{r},t)$, the total power dissipated per unit volume is given by:
\begin{equation}
   \dot{Q}(\vec{r},t) = 2 \kappa_\text{loc}(\vec{r},t) \varepsilon_0 \varepsilon_r(\vec{r}) \vec{e}(\vec{r}) \cdot \vec{e}^{\,*}(\vec{r})|a(t)|^2,
   \label{eq:heat-source}
\end{equation}
where $|a(t)|^2$ is the number of photons circulating in the optical cavity, $\vec{e}(\vec{r})$ is the electromagnetic mode profile, $\varepsilon_0$ and $\varepsilon_r$ are the vacuum permittivity and dielectric function, respectively. 


We would like to express Eq.~\ref{theta-evo-int} in terms of the experimentally accessible total absorptive loss rate $\kappa_\text{abs}(t)$ (the time dependency arises if nonlinear absorption is considered) instead of $\kappa_{\text{loc}}(\vec{r},t)$. These quantities are related by:

\begin{equation}
    \kappa_\text{abs}(t) = \frac{\int \kappa_{\text{loc}}(\vec{r},t)\varepsilon_r(\vec{r}) |\vec{e}(\vec{r})|^2 dV  }{\int \varepsilon_r(\vec{r}) |\vec{e}(\vec{r})|^2 dV}.
\end{equation}

Since $\kappa_\text{abs}(t)$ is part of a lumped model, it is clear that with its knowledge alone, one cannot solve for $\kappa_{\text{loc}}(\vec{r},t)$. Instead, if linear absorption is considered, we simplify matters by assuming $\kappa_{\text{loc}}(\vec{r},t)$ is a constant $\kappa_{\text{loc,}0}$ within an absorbing volume $V_\text{abs}$. In that case, we may solve for $\kappa_{\text{loc,}0}$ in terms of the total linear absorption $\kappa_\text{abs} = \eta_0 \kappa_0$, where $\kappa_0$ denotes the cold-cavity (linear) intrinsic losses of the device, and $\eta_0$ its absorptive fraction. We have:

\begin{equation}
    \kappa_{\text{loc},0} = \eta_0 \kappa_{0}\frac{\int \varepsilon_r(\vec{r}) |\vec{e}(\vec{r})|^2 dV  }{\int_{V_\text{abs}} \varepsilon_r(\vec{r}) |\vec{e}(\vec{r})|^2 dV}.
\end{equation}

Although assuming homogeneous absorption within the absorbing volume is certainly an approximation, it is generally sufficient to correctly capture optical absorption even when it takes place near the surface. For example, in case the main mechanism of absorption is mediated by surface imperfections, we should expect an exponential decay in $\kappa_{\text{loc}}(\vec{r})$ as we move away from the surface. In that case, we could simply treat $\kappa_{\text{loc}}$ and $V_\text{abs}$ as effective quantities.

In order to incorporate nonlinear absorption mechanisms such as two-photon (TPA) and free-carrier (FCA) absorption, the spatio-temporal dependency of $\kappa_{\text{loc}}(\vec{r},t)$ must be related to the fields in the resonator. For phenomena that scale linearly with the electromagnetic energy in the resonator, e.g. TPA~\cite{Barclay2005NonlinearTaper}, we expect $\kappa_{\text{loc}}(\vec{r},t)$ to have the following form:
\begin{equation}
   \kappa_{\text{loc,}1}(\vec{r},t) =  C_{1} \frac{\varepsilon_r(\vec{r}) |\vec{e}(\vec{r})|^2}{\int \varepsilon_r(\vec{r}) |\vec{e}(\vec{r})|^2 dV} \times |a(t)|^2, 
\end{equation}
since the fields themselves generate absorption. Analogously, for phenomena that scale quadratically with the electromagnetic energy in the resonator, e.g FCA:

\begin{equation}
   \kappa_{\text{loc,}2}(\vec{r},t) =  C_\text{2}\frac{\varepsilon_r(\vec{r})^2 |\vec{e}(\vec{r})|^4}{(\int \varepsilon_r(\vec{r}) |\vec{e}(\vec{r})|^2 dV)^2} \times |a(t)|^4, 
\end{equation}
where nonlocal effects such as carrier diffusion are neglected.

In both cases above, $C_{2}$ and $C_{1}$ are approximated as material-dependent constants which are only appreciable inside the volume of the cavity, defined by $V_\text{cav}$. They can be expressed in terms of the experimentally accessible quantities such as the first ($\kappa_{\text{abs},2}(t)$) and second ($\kappa_{\text{abs},2}(t)$) order absorption rates:

\begin{equation}
    C_\text{1} = \frac{\kappa_{1}(t)}{|a(t)|^2} \frac{( \int \varepsilon_r(\vec{r}) |\vec{e}(\vec{r})|^2 dV )^2 }{\int_{V_\text{cav}} \varepsilon_r^2(\vec{r}) |\vec{e}(\vec{r})|^4 dV},
\end{equation}

\begin{equation}
    C_\text{2} = \frac{\kappa_{2}(t)}{|a(t)|^4} \frac{( \int \varepsilon_r(\vec{r}) |\vec{e}(\vec{r})|^2 dV )^3 }{\int_{V_\text{cav}} \varepsilon_r^3(\vec{r}) |\vec{e}(\vec{r})|^6 dV}.
\end{equation}

Writing $\kappa_\text{loc}(\vec{r},t) = \kappa_{\text{loc,}0}+\kappa_{\text{loc,}1}(\vec{r},t)+\kappa_{\text{loc,}2}(\vec{r},t)$ in Eq.~\ref{eq:heat-source} and using the definitions of $C_\text{1}$ and $C_\text{2}$ above,  equation \ref{theta-evo-int} can be written in terms of the photon energy $\hbar \omega_l$ and the zeroth, first, and second order thermal resistances, yielding:

\begin{equation}
    \dot{\theta}_k = -\frac{1}{\tau_k}\theta_k +  \frac{\hbar \omega_l }{\tau_k}(\eta_0\kappa_{0} R^\theta_{k,0} + \kappa_{1}(t) R^\theta_{k,1} + \kappa_{2}(t) R^\theta_{k,2})\times |a(t)|^2,
    \label{nl_theta-evo}
\end{equation}
with:
\begin{equation}
    R^\theta_{k, 0} =\frac{2 \tau_k \varepsilon_0}{\hbar \omega_l} \frac{ \int_{V_\text{abs}}{\varepsilon_r (\vec{r}) |\vec{e}(\vec{r})|^2\delta \Tilde T_k(\vec{r})dV}}{\int{c_p\rho (\delta \Tilde T_k)^2  dV}} \frac{\int \varepsilon_r(\vec{r}) |\vec{e}(\vec{r})|^2 dV  }{\int_{V_\text{abs}} \varepsilon_r(\vec{r}) |\vec{e}(\vec{r})|^2 dV},
    \label{thermal_resistance}
\end{equation}

\begin{equation}
    R^\theta_{k, 1} =\frac{2 \tau_k \varepsilon_0}{\hbar \omega_l} \frac{ \int_{V_\text{cav}}{\varepsilon_r^2 (\vec{r}) |\vec{e}(\vec{r})|^4\delta \Tilde T_k(\vec{r})dV}}{\int{c_p\rho (\delta \Tilde T_k)^2  dV}}  \frac{\int \varepsilon_r(\vec{r}) |\vec{e}(\vec{r})|^2 dV  }{\int_{V_\text{cav}} \varepsilon_r^2(\vec{r}) |\vec{e}(\vec{r})|^4 dV},
    \label{thermal_resistance_TPA}
\end{equation}

\begin{equation}
    R^\theta_{k, 2} =\frac{2 \tau_k \varepsilon_0}{\hbar \omega_l} \frac{ \int_{V_\text{cav}}{\varepsilon_r^3 (\vec{r}) |\vec{e}(\vec{r})|^6\delta \Tilde T_k(\vec{r})dV}}{\int{c_p\rho (\delta \Tilde T_k)^2  dV}}  \frac{\int \varepsilon_r(\vec{r}) |\vec{e}(\vec{r})|^2 dV  }{\int_{V_\text{cav}} \varepsilon_r^3(\vec{r}) |\vec{e}(\vec{r})|^6 dV}.
    \label{thermal_resistance_FCA}
\end{equation}

Eqs.~\ref{thermal_resistance}, \ref{thermal_resistance_TPA}, \ref{thermal_resistance_FCA} can be evaluated in a straightforward way by any electromagnetic/thermal solver. Mechanical vibrations may also lead to heating through thermoelastic damping, however, due to its negligible dynamical contributions in the devices discussed here, this effect will be neglected. Importantly, if $\delta\tilde{T}_k(\vec{r})$ is approximately constant in the region of overlap with the optical mode, all three expressions above yield similar values. This approximation is well justified for the first few low-order thermal modes, which turn out to be the most relevant for the case treated in the main text.

In practice, not all of the nonlinear absorptive losses ($\kappa_1$ and $\kappa_2$) are necessarily converted to heat, as implied by Eq.~\ref{nl_theta-evo}. Nonlinear absorption is strongly tied to carrier dynamics inside the device. As we are dealing with a direct band-gap semiconductor, we expect that carriers generated through virtual processes such as TPA to display both radiative and non-radiative recombination routes. That is, only a fraction of the nonlinear loss rate acts as a heat source in our resonator. As a consequence, we are led to define the ratios $\eta_i$ of the $i$th order nonlinear loss rate that contributes to the heating of the resonator. With that in mind, the generalized evolution of the amplitude $\theta_k(t)$ is given by:

\begin{equation}
    \dot{\theta}_k = -\frac{1}{\tau_k}\theta_k +  \frac{\hbar \omega_l }{\tau_k}(\eta_0 \kappa_{0} R^\theta_{k,0} + \eta_1 \kappa_{1}(t) R^\theta_{k,1} + \eta_2\kappa_{2}(t) R^\theta_{k,2})\times |a(t)|^2,
\end{equation}

\section{Frequency domain description of the photothermal coupling}\label{sec:freq_domain}

\subsection{Single-mode Photothermal coupling}\label{subsec:freq_domain_1}

We start from the coupled equations for the temporal evolution of the optical, mechanical and temperature fields. In this subsection, we assume that the dynamics is dominated by one mode in each of the previously cited domains, therefore, we drop all mode-related indices. For simplicity, in this and in the next subsection we treat the case of linear absorption, i.e. $\kappa_{\text{abs},1}(t)=\kappa_{\text{abs},2}(t) = 0$. The equations read:
\begin{equation}
\begin{split}
    \dot{a} = i\Delta(t) a -\frac{\kappa}{2} a +\sqrt{\kappa_e}\alpha_{in},\\
    m_\text{eff}(\ddot{x}+\Gamma\dot{x}+\Omega^2 x) = F^L(t)+F^\text{RP}(t)+F^\theta(t),\\
    \dot{\theta} = -\frac{1}{\tau}\theta + \frac{\hbar \omega_l \kappa_\text{abs}R^\theta}{\tau}|a|^2, 
\end{split}
\label{coupled_equations}
\end{equation}
where $\Delta(t) = \omega_l-\omega_c +G^{\theta} \theta+G^x x$, is the relative detuning between the cavity and the laser, including optomechanical and thermal frequency pullings, $\omega_c$ is the (optical) resonance frequency and $\omega_l$ the external laser frequency. $F^\text{L}$, $F^\text{RP}$ and $F^{\theta}$ are respectively the thermal Langevin, the radiation pressure and the photothermal (bolometric) forces, $G^x$ and $G^{\theta}$ are respectively the optomechanical and thermo-optical coupling rates. We also choose a normalization for the electromagnetic modes such that $|a|^2$ is the number of circulating photons in the cavity, consequently, $\alpha_{in}$ is the rate of incident photons on the resonator, driven by an external source.

We may derive expressions for $G^{\theta}$ and $G^x$ using first-order perturbation theory on Maxwell's equations~\cite{Johnson2002PerturbationBoundaries, Balram2014MovingResonators}:

\begin{subequations}
\begin{gather}
    G^{\theta} = \omega_c \frac{\int{|~\vec{e}~|^2 n \frac{dn}{dT}\delta \Tilde T dV}}{\int{|~\vec{e}~|^2 \varepsilon dV}}
    \label{eq:pert_theorya},\\
    G^x = -\frac{\omega_c}{2} \frac{\int{\vec{e} \cdot (\bm p \mkern1mu{:} \bm S^x) \cdot \vec{e}\,^* dV}}{\int{|~\vec{e}~|^2 \varepsilon dV}}  + \frac{\omega_{c}}{2}\frac{\int d\vec{A}{\cdot}\vec{u}\left(\Delta\bm\varepsilon|\vec e_{\parallel}|^2-\Delta\bm\varepsilon^{-1}|\vec d_{\perp}|^2 \right)}{\int |~\vec{e}~|^2 \varepsilon dV},
    \label{eq:pert_theoryb}
\end{gather}
\end{subequations}
where  Eq.~\ref{eq:pert_theorya} is related to the thermo-optic effect, related to temperature induced changes in the dielectric response of the medium. The integrals in Eq.~\ref{eq:pert_theoryb} model the photoelastic and moving boundary optomechanical couplings related to a mechanical mode $\vec{u}$, where $\bm p$ is the photoelastic tensor.  The moving boundary contribution is a function of $\Delta\bm\varepsilon=\bm\varepsilon_1-\bm\varepsilon_2$ and $\Delta\bm\varepsilon^{-1}~=~(\bm\varepsilon_1)^{-1}-(\bm\varepsilon_2)^{-1}$, which are related to the permittivities of the guiding ($\bm\varepsilon_1$) and surrounding ($\bm\varepsilon_2$) materials. 

In the spirit of deriving an entirely linear theory, we look for linearized solutions to Eqs.~\ref{coupled_equations} as $a(t)=a_0+\delta a(t)$, $x(t) = x_0 +\delta x(t)$, and $\theta(t) = \theta_0 +\delta\theta(t)$. Using $F^\text{RP} = \hbar G^x |a|^2$, $F^{\theta} = \Lambda^{\theta} \theta$ and collecting the fluctuating terms up to first order we obtain the following dynamical equations for thermo-optomechanics (initially disregarding the Langevin term):

\begin{equation}
\begin{split}
  \delta\dot{a}=i(
  G^{x}\delta x+G^{\theta}\delta\theta)a_{0}+(i\Delta_0 -\frac{\kappa}{2})\delta a,\\
  \delta\ddot{x}+\Gamma\delta\dot{x}+\Omega^{2}\delta x = \frac{\hbar G^{x}}{m_\text{eff}}(a_0^{*}\delta a+ a_0\delta a^{*})+\frac{\Lambda^\theta}{m_\text{eff}} \delta\theta,\\
 \delta \dot{\theta} = -\frac{1}{\tau}\delta\theta + \frac{\hbar \omega_l \kappa_\text{abs}R^\theta}{\tau}(a_0^{*}\delta a+ a_0\delta a^{*}),
\end{split}
\end{equation}
where $\Delta_0$ absorbs the static frequency shifts given by $G^{\theta} \theta_0+G^x x_0$. In frequency space:
\begin{equation}
\begin{split}
  [(\Delta_0+ \omega) +i\frac{\kappa}{2}]\delta a(\omega)= -\big[ G^{x}\delta x(\omega)+G^{\theta}\delta\theta(\omega) \big] a_{0},\\
  \big[(\Omega^{2}-\omega^2)-i\omega \Gamma\big] \delta x(\omega)= \frac{\hbar G^{x}}{m_\text{eff}}\big[a_0^{*}\delta a(\omega)+ a_0[\delta a]^*(\omega)\big]+\frac{\Lambda^\theta}{m_\text{eff}} \delta\theta(\omega),\\
  -i\omega \delta\theta(\omega) = -\frac{1}{\tau}\delta\theta(\omega) + \frac{\hbar \omega_l \kappa_\text{abs}R^\theta}{\tau}\big[a_0^{*}\delta a(\omega)+ a_0[\delta a]^*(\omega)\big],
\end{split}
\end{equation}
since $[\delta a]^* (\omega) = [\delta a(-\omega)]^*$ and for real variables such as $\delta \theta$ and $\delta x$, $\delta x(\omega) = [\delta x(-\omega)]^*$, we may eliminate $\delta a(\omega)$ and $[\delta a]^*(\omega)$, obtaining:
\begin{subequations}
\begin{align}
       \delta\theta(\omega) = - \frac{\hbar \omega_l \kappa_\text{abs}R^\theta\chi^\theta(\omega)\Psi(\omega,\Delta_0)|a_0|^2 G^x}{\big[\tau + \hbar \omega_l \kappa_\text{abs}R^\theta G^{\theta}\chi^{\theta}(\omega)\Psi(\omega,\Delta_0)|a_0|^2\big]} \delta x (\omega),\\
       \big[(\Omega^{2}-\omega^2)-i\omega \Gamma\big] \delta x(\omega)= -\frac{\hbar G^{x}}{m_\text{eff}}\Psi(\omega,\Delta_0)|a_0|^2(G^{\theta}\delta\theta(\omega)+ G^x \delta x(\omega))+\frac{\Lambda^\theta}{m_\text{eff}} \delta\theta(\omega),
       \label{eq:fourier_a_theta}
\end{align}
\end{subequations}
where $\Psi(\omega,\Delta_0) = [\frac{1}{(\Delta_0+\omega)+i\kappa/2}+\frac{1}{(\Delta_0-\omega)-i\kappa/2}]$, a combination of the cavity's optical susceptibility and its conjugate, and  $\chi^\theta(\omega) = \frac{1}{1/\tau -i\omega}$ is the thermal susceptibility.

Eliminating $\delta\theta(\omega)$ in Eq.~\ref{eq:fourier_a_theta}, we readily obtain the dressed optomechanical self-energy. The term on the RHS which is being multiplied by $G^{\theta}$ yields the correction acting on the radiation pressure feedback, therefore the total $RP$ modification to the linear response of the system is:
\begin{equation}
    \Sigma_\text{eff}^\text{RP}(\omega) = \frac{\Sigma^\text{RP}(\omega)}{\big[1 + \frac{\hbar \omega_l \kappa_\text{abs}R^\theta G^{\theta}\chi^{\theta}(\omega)\Psi(\omega,\Delta) |a_0|^2}{\tau}\big]},
\end{equation}
where we defined $\Sigma^\text{RP}(\omega) = \hbar (G^x)^2 \Psi(\omega,\Delta_0) |a_0|^2$, the optically induced mechanical inverse susceptibility in the absence of thermal feedback. 
The correction above is negligible for the GaAs microdisks considered in the main text, however it might become relevant in other systems where mechanical frequencies are comparable to or smaller than the inverse of thermal relaxation times. In that case, thermal dispersion affects the optical spectrum within the mechanical timescale, impacting optomechanical transduction.

The term proportional to $\Lambda^\theta$ in Eq.~\ref{eq:fourier_a_theta} yields the dressed bolometric force, which is this work's object of study. The photothermal contribution to the inverse mechanical susceptibility is given by:

\begin{equation}
    \Sigma^\theta_{\text{eff}}(\omega) =  \frac{\Sigma^\theta(\omega)}{\big[1 + \frac{\hbar \omega_l \kappa_\text{abs}R^\theta G^{\theta}\chi^{\theta}(\omega)\Psi(\omega,\Delta) |a_0|^2}{\tau}\big]},
    \label{bolometric}
\end{equation}
where we defined the bare bolometric contribution to the inverse mechanical susceptibility as:

\begin{equation}
    \Sigma^\theta(\omega) =  \frac{ \hbar \omega_l \kappa_\text{abs}R^\theta \Lambda^\theta G^x \chi^\theta(\omega)\Psi(\omega,\Delta)|a_0|^2}{\tau}.
\end{equation}

The dressed inverse mechanical susceptibility $\chi_{m, \mathrm{eff}}^{-1}(\omega)$, including RP and photothermal contributions is finally given by:

\begin{equation}
    \chi_{m, \mathrm{eff}}^{-1}(\omega)=\chi_{m}^{-1}(\omega)+\Sigma_\text{eff}^\text{RP}(\omega)+\Sigma^\theta_{\text{eff}}(\omega),
\end{equation}
where $\chi_{m}^{-1}(\omega)=m_{\mathrm{eff}}\left[\left(\Omega^{2}-\omega^{2}\right)-i \Gamma \omega\right]$ is the inverse of the bare mechanical susceptibility.

The expressions above for $\Sigma^{\text{RP}}_{\text{eff}}$ and $\Sigma^{\theta}_{\text{eff}}$ may be simplified in many practical scenarios. In agreement with the experimental case reported in the main text, we assume the bad-cavity regime ($\Omega \ll \kappa/2$) and $\tau \gg 1/\Omega$. Evaluating all frequency responses at $\Omega$ , which is the driving frequency given by the mechanical oscillations, and also taking the detuning $\Delta_0 = \kappa/2$, we note that $\chi^\theta \approx i/\Omega$, and $\Psi \approx 2/\kappa$. For typical microdisks, $\tau/(2\pi) \approx 10$ $\mu$s, $\Omega/(2\pi) \approx 250$ MHz,  $\kappa_\text{abs}/(2\pi) \approx 1$ GHz, $\omega_l/(2\pi)\approx 200$ THz, $R^\theta \approx 4\times 10^4$, $G^{\theta}/(2\pi) \approx 10$ GHz, $\kappa/(2\pi) \approx 10$ GHz and $|a_0|^2 \approx 10^4$. With these numbers we see that: 

\begin{equation}
    \frac{\hbar \omega_l \kappa_\text{abs}R^\theta G^{\theta}\chi^{\theta}(\omega)\Psi(\omega,\Delta) |a_0|^2}{\tau} << 1.
\end{equation}
Confirming that the correction due to the thermo-optical dispersion is negligible in our case.

One last approximation can be made in order to compare $\Sigma^\theta$ and $\Sigma^\text{RP}$. Considering only cooling/amplification under the same approximations as before, the ratio of photothermal ($\delta\Gamma^\theta$) and RP ($\delta\Gamma^\text{RP}$) mechanical amplification, evaluated at $\Delta_0 = \kappa/2$ is given by:

\begin{equation}
    \frac{\delta\Gamma^\theta}{\delta\Gamma^\text{RP}}(\Delta = \kappa/2) = \frac{1}{2}\frac{\omega_l \kappa_\text{abs} R^\theta \Lambda^\theta \kappa}{\tau\Omega^2 G^x},
    \label{ratio_simplifyed}
\end{equation}
which can be used as a simple test for the relevance of photothermal forces in devices that operate within the regimes above.

\subsection{Multimode Photothermal coupling}\label{subsec:freq_domain_2}

We now generalize our discussion to the case where several thermal modes couple relevantly to optical and mechanical modes. The appropriate equations in frequency space are given by:

\begin{equation}
\begin{split}
  [(\Delta_0+ \omega) +i\frac{\kappa}{2}]\delta a(\omega)= -\big[ G^{x}\delta x(\omega)+\sum_k G^{\theta}_k \delta\theta_k(\omega) \big] a_{0},\\
  \big[(\Omega^{2}-\omega^2)-i\omega \Gamma\big] \delta x(\omega)= \frac{\hbar G^{x}}{m_\text{eff}}\big[a_0^{*}\delta a(\omega)+ a_0[\delta a]^*(\omega)\big]+\sum_k\frac{\Lambda^{\theta}_k}{m_\text{eff}} \delta\theta_k(\omega),\\
  -i\omega \delta\theta_k(\omega) = -\frac{1}{\tau_k}\delta\theta_k(\omega) + \frac{\hbar \omega_l \kappa_\text{abs}R^\theta_k}{\tau_k}\big[a_0^{*}\delta a(\omega)+ a_0[\delta a]^*(\omega)\big],\,k = 1,2,3....
\end{split}
\label{eq:dyn}
\end{equation}

Eliminating $\delta a(\omega)$ from the equations for the thermal modes we get:

\begin{equation}
\delta\theta_k(\omega) = \frac{A^{\theta}_k}{(1-A^{\theta}_kG^{\theta}_k)}\big(G^x \delta x (\omega)+ \sum_{l\neq k} G^{\theta}_l \delta\theta_l(\omega)\big),
\end{equation}
where we defined:

\begin{equation}
    A^{\theta}_k(\omega,\Delta_0) = - \frac{\hbar \omega_l \kappa_\text{abs}R^\theta_k\chi^{\theta}_k (\omega)\Psi(\omega,\Delta_0)|a_0|^2}{\tau_k}.
\end{equation}
By inspection, one may show that $\frac{\delta\theta_l(\omega)}{A^{\theta}_l(\omega,\Delta_0)}=\frac{\delta\theta_k(\omega)}{A^{\theta}_k(\omega,\Delta_0)}$. With that result we get:

\begin{equation}
    \delta \theta_k(\omega) = \frac{A^{\theta}_k(\omega,\Delta_0) G^x}{1-\sum_k A^{\theta}_k(\omega,\Delta_0) G^{\theta}_k }\delta x (\omega),
\end{equation}
and the total photothermal modification to the inverse mechanical susceptibility is:

\begin{equation}
    \Sigma^{\theta}_{\text{eff}}(\omega) =  \frac{\Sigma^\theta}{1-\sum_k A^{\theta}_k(\omega,\Delta_0) G^{\theta}_k },
\label{eq:suscept_bolo_multi}
\end{equation}
where the bare photothermal contribution is given by:

\begin{equation}
    \Sigma^\theta(\omega) =  \sum_k A^{\theta}_k(\omega,\Delta_0) \Lambda^{\theta}_k G^x,
\end{equation}

In analogy with the single thermal mode case, the corrected RP induced inverse susceptibility is given by:


\begin{equation}
    \Sigma^\text{RP}_{\text{eff}}(\omega) = \frac{\Sigma^\text{RP}(\omega)}{1-\sum_k A^{\theta}_k(\omega,\Delta_0) G^{\theta}_k }.
    \label{eq:suscept_RP_multi}
\end{equation}

From the symmetries of the equations above, it is useful to define ``effective thermal response'' functions, at a given frequency $\omega$:

\begin{equation}
    \begin{split}
        h_{1}^\theta(\omega) = \kappa_\text{abs} \sum_k \frac{R^\theta_k\Lambda^{\theta}_k\chi^{\theta}_k (\omega)}{\tau_k},\\
        h_{2}^\theta(\omega) = \kappa_\text{abs} \sum_k \frac{R^\theta_k G^{\theta}_k \chi^{\theta}_k(\omega)}{\tau_k},
    \end{split}
    \label{eq:therm_resp_funcs}
\end{equation}
in terms of which, expressions \ref{eq:suscept_bolo_multi} and \ref{eq:suscept_RP_multi} can be recast as:

\begin{equation}
    \begin{split}
        \Sigma^\theta_{\text{eff}}(\omega) =  \frac{\hbar \omega_l \Psi(\omega,\Delta_0)|a_0|^2 G^x h_{1}^\theta(\omega)}{1+\hbar \omega_l \Psi(\omega,\Delta_0)|a_0|^2h_{2}^\theta(\omega)},\\
        \Sigma^\text{RP}_{\text{eff}}(\omega) =  \frac{\hbar \Psi(\omega,\Delta_0)|a_0|^2 (G^x)^2}{1+\hbar \omega_l \Psi(\omega,\Delta_0)|a_0|^2h_{2}^\theta(\omega)},
        \label{eq:multi_all}
    \end{split}
\end{equation}
which share the same form as their single mode counterparts. We note that the summations for $h_{1}^\theta(\omega)$ and $h_{2}^\theta(\omega)$ are well behaved since $R^\theta_k/\tau_k$ is independent of $\tau_k$ and $\chi^{\theta}_k$ is a monotonically decreasing function of $\tau_k$, guaranteeing gradually decreasing importance of high-order thermal modes ($\tau_k \to 0$).

From the equations above it is straightforward to compare the photothermal and radiation pressure forces amplitudes. The ratio $\Sigma^\theta/\Sigma^\text{RP}$ yields:

\begin{equation}
    \frac{\Sigma^\theta_{\text{eff}} (\omega)}{\Sigma^\text{RP}_{\text{eff}}(\omega)} = \frac{\hbar \omega_l h_{1}^\theta(\omega)}{\hbar G^x}.
\end{equation}
Since the denominator of the RHS essentially gives the RP force per photon, the numerator may be interpreted likewise, but regarding the photothermal force. Notice however, that due to non-zero relaxation times, the two forces display a non-trivial phase relation, here captured by the complex part of $h^\theta_{1} (\omega)$. 

The approximation obtained in the previous section can be generalized to the multimode case. Again, considering the bad-cavity limit and neglecting thermal dispersion, i.e. $h^\theta_{2}(\Omega)\ll 1$:

\begin{equation}
    \frac{\delta\Gamma^\theta}{\delta\Gamma^\text{RP}}(\Delta = \kappa/2) = \frac{1}{2}\frac{\omega_l \kappa \text{Im}[h_{1}^\theta]}{\Omega G^x},
    \label{multimode_PT_relevance}
\end{equation}

\subsection{Nonlinear Multimode Photothermal coupling}\label{subsec:freq_domain_3}
In this subsection, we summarize the main results concerning the usage of Eq.~\ref{eq:dyn}, without neglecting the nonlinear losses in general. That is, we use Eq.~\ref{nl_theta-evo} and make $\kappa\rightarrow \kappa(|a|^2)$ in the evolution of $a$. That is, the total nonlinear intrinsic loss, $\kappa_i(t)$, is given by $\kappa_i(t) = \kappa_{0}+\kappa_{1}(t)+ \kappa_{2}(t)$. The coupled equations of the optical, mechanical and thermal responses read:

\begin{equation}
\begin{split}
    \dot{a} = i\Delta(t) a -\frac{\kappa_e +  \kappa_{0}+\kappa_{1}(t)+ \kappa_{2}(t)}{2} a +\sqrt{\kappa_e}\alpha_{in},\\
    m_\text{eff}(\ddot{x}+\Gamma\dot{x}+\Omega^2 x) = F^L(t)+F^\text{RP}(t)+F^\theta(t),\\
     \dot{\theta}_k = -\frac{1}{\tau_k}\theta_k +  \frac{\hbar \omega_l }{\tau_k}(\eta_0 \kappa_{0} R^\theta_{k,0} + \eta_1\kappa_{1}(t) R^\theta_{k,1} + \eta_2\kappa_{2}(t) R^\theta_{k,2})\times |a(t)|^2,\,k = 1,2,3..., 
\end{split}
\label{eq:nl_dyn}
\end{equation}
where $\Delta(t) = \omega_l-\omega_c +\sum_k G_k^\theta \theta(t)+G^x x(t)$.

A fundamental difference between the present case and the analysis in past sections is that, when considering the frequency domain version of the above equations, the Fourier transform of the nonlinear losses must also be performed, since those also display a time dependency, substantially modifying the final results.

In our model for the nonlinear losses we neglected all but the dynamics of the energy in the resonator. That is, the only time dependency of the first and second order dissipation is given through the number of photons $|a(t)|^2$. Since those losses are intrinsically related to carrier dynamics, this is explicitly an approximation. Its validity requires that any timescale associated to carrier relaxation (i.e. its recombination rate) is must faster than the period of mechanical oscillations, in such a way that we may assume that their population responds instantly to energy variations. As stated in past work on this subject, typical recombination rates for GaAs range around $\SI{10}{\ps}$~\cite{Guha2017Surface-enhanced106}, much faster than the $\approx \SI{4}{\ns}$ associated with mechanical oscillations.

Under the aforementioned hypothesis, we write $\kappa_1(t) = \kappa'_1 |a(t)|^2$ and $\kappa_2(t) = \kappa'_2 |a(t)|^4$, where the $\kappa'_i$ are constants. This form allows us to linearize the losses according to $a(t)\rightarrow a_0+\delta a(t)$. After linearizing and writing Eqs.~\ref{eq:nl_dyn} in frequency space, they read:

\begin{equation}
\begin{split}
    \left[i\Delta_0-\frac{\kappa_0+2\bar{\kappa}_1+3\bar{\kappa}_2}{2}\right]\delta a(\omega )-\frac{a_0}{a_0^*}\left[\frac{\bar{\kappa}_1+2\bar{\kappa}_2}{2}\right][\delta a]^*(\omega ) = -i [\sum_k G^{\theta}_k \delta \theta_k (\omega )+a_0 G^x \delta
   x(\omega )] a_0,\\
    \big[(\Omega^{2}-\omega^2)-i\omega \Gamma\big] \delta x(\omega)= \frac{\hbar G^{x}}{m_\text{eff}}\big[a_0^{*}\delta a(\omega)+ a_0[\delta a]^*(\omega)\big]+\sum_k\frac{\Lambda^{\theta}_k}{m_\text{eff}} \delta\theta_k(\omega),\\
     -i\omega\theta_k(\omega) = -\frac{1}{\tau_k}\theta_k(\omega) +  \frac{\hbar \omega_l }{\tau_k}(\eta_0\kappa_{0} R^\theta_{k,0} + 2\eta_1\bar{\kappa}_1 R^\theta_{k,1} + 3\eta_2\bar{\kappa}_2 R^\theta_{k,2})\times \left(a_0[\delta a]^* (\omega)+a_0^*\delta a (\omega)\right),\,k = 1,2,3..., 
\end{split}
\label{eq:nl_four}
\end{equation}
where we defined the average losses $\bar{\kappa}_j = \kappa'_j|a_0|^{2j}$ and $\bar{\kappa}_0 = \kappa_0$

We follow the same procedure as before: we eliminate $\theta(\omega)$ and $\delta a(\omega)$ from the equation for the mechanical response, from which we recover a dressed susceptibility for the mechanical mode. The PT-induced and the RP-induced modifications to the mechanical susceptibility are:

\begin{eqnarray}
    \Sigma^{\theta}_{\text{eff}}(\omega) &=&  \frac{\hbar \omega_l\Psi_{\text{dyn}}(\omega,\Delta_0)|a_0|^2 G^x h_{1,\text{dyn}}^\theta(\omega)}{1+\hbar \omega_l \Psi_{\text{dyn}}(\omega,\Delta_0) |a_0|^2 h_{2,\text{dyn}}^\theta(\omega)}\\
    \Sigma^\text{RP}_{\text{eff}}(\omega) &=&  \frac{\hbar \Psi_{\text{dyn}}(\omega,\Delta_0)|a_0|^2(G^x)^2}{1+\hbar \omega_l \Psi_{\text{dyn}}(\omega,\Delta_0) |a_0|^2h_{2,\text{dyn}}^\theta(\omega)},
\end{eqnarray}
where $h_{1,\text{dyn}}^\theta(\omega)$ and $h_{2,\text{dyn}}^\theta(\omega)$ are, respectively, the effective responses for the photothermal force and for the thermo-optic dispersion:
\begin{equation}
    \begin{split}
        h_{1,\text{dyn}}^\theta(\omega) = \sum^2_{j=0} (j+1) \eta_j \bar{\kappa}_j \sum_k \frac{R^\theta_{k,j}\Lambda^{\theta}_k\chi^{\theta}_k (\omega)}{\tau_k},\\
        h_{2,\text{dyn}}^\theta(\omega) = \sum^2_{j=0} (j+1) \eta_j \bar{\kappa}_j \sum_k \frac{R^\theta_{k,j} G^{\theta}_k \chi^{\theta}_k(\omega)}{\tau_k},
    \end{split}
    \label{eq:nl_therm_resp_funcs}
\end{equation}
and $\Psi_{\text{dyn}}$ is the modified optical response due the dynamic of the nonlinear losses:

\begin{equation}
    \Psi_{\text{dyn}}(\Delta_0,\omega) = \frac{8\bar{\Delta}_0}{4\Delta_0^2 + \Big(\sum^2_{j=0}(2j+1)\bar{\kappa}_j- 2 i \omega \Big)\Big(\sum^2_{j=0}\bar{\kappa}_j -2 i \omega \Big)}.
\end{equation}


The equations above were used in the analysis of the experiment reported in the main text.

\section{Localized Absorption}\label{sec:absorption}
\begin{figure*}[ht!]
\centering
\includegraphics[width = 16.0cm]{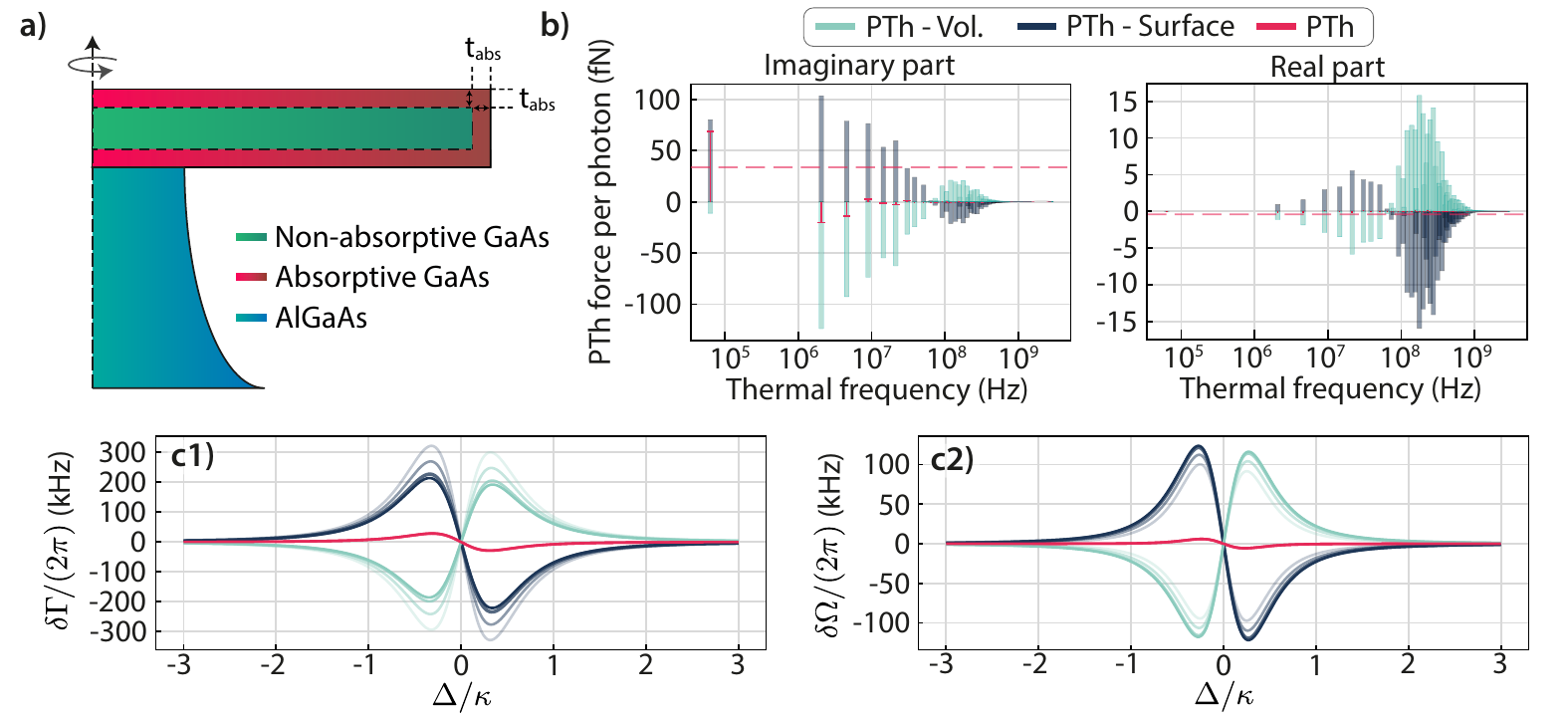}
\caption{\small{\textbf{(a)} Schematic of absorbing and non-absorbing regions of the GaAs microdisk. The variable $t_\text{abs}$ defines the thickness of the absorbing layer. \textbf{(b)} PTh force per photon (real and imaginary components) for each of the first 200 thermal modes, here separated by frequency. \textbf{(c1)} Mechanical linewidth and \textbf{(c2)} frequency variations as a function of the normalized detuning $\Delta/\kappa$ for different $t_\text{abs}$. From lighter to darker colors: $t_\text{abs}= \SI{15}{}$,  $\SI{62.5}{}$, $\SI{110}{}$, and $\SI{125}{\nm}$ (entire GaAs membrane).}}
\label{supp:SX}
\end{figure*}

In this section we use FEM simulations to investigate how surface localized absorption affects the photothermal effects on our microdisk devices. An absorptive layer with thickness $t_{\text{abs}}$ is defined within the GaAs membrane composing the microdisk illustrated in Fig.~\ref{supp:SX} \textbf{a)}. Importantly, the absorption is taken to be constant over this layer. The radius of the disk is $\SI{6}{\micro\metre}$ and the pedestal diameter is $\SI{1.5}{\micro\metre}$, the latter is measured in the region of contact of the AlGaAs/GaAs layers. The geometry of this device agrees with the one used in Fig. 2 in the main text. In the following analysis we consider the first order TE optical mode and the first order mechanical breathing ($\Omega/(2\pi) = \SI{230}{\MHz}$) mode of the microdisk. 

In the main text we study the case of an homogeneously absorbing GaAs membrane (no absorption layer), in which the total PTh force per mode tends to become negligibly small for higher order thermal modes ($k>5$), this happens due to the self-cancellation of the surface and volume contributions. This is shown in Fig.~\ref{supp:SX} \textbf{b)} where each thermal mode's contribution to the imaginary and real parts total PTh force per photon was evaluated. As the thermal frequency -- defined as $\tau^{-1}/(2\pi)$ -- increases, the red T-shaped bars in the imaginary part of such force become negligibly small. Nevertheless, we verify appreciable -- contrary -- surface and volume forces for frequencies as high as $\SI{1}{\GHz}$, in agreement with the discussion in the main text. The total real component of the force (horizontal dashed line) is evaluated to be approximately two orders of magnitude smaller than its quadrature and its related red bars are barely seen when compared to the dark blue and cyan bars.

In Fig.~\ref{supp:SX} \textbf{c1)} and \textbf{c2)} we evaluate the consequences of PTh backaction on the mechanical mode's linewidth and frequency, respectively, for several different absorption layers, with $t_\text{abs} =  \SI{15},\,\SI{62.5},\,\SI{110}{\nm}$ and also for the case of homogeneously distributed absorption in the GaAs membrane. The total surface and volume contributions are appreciably altered by the absorptive layer thickness, remarkably the overall response is essentially unaffected. This is understood under the light of the results of Fig.~\ref{supp:SX} \textbf{b)}. Since only low-order thermal modes are important for the device's overall response and those are essentially homogeneous over the electromagnetic mode's confinement region, their response is unaffected due to the presence of an absorption layer. As the surface and volume forces are still appreciable for high-order modes -- and those display rapid spatial oscillations --, their response is significantly modified by localized absorption. We conclude that the differences verified in the navy/cyan curves in Fig.~\ref{supp:SX} \textbf{c1)} and \textbf{c2)} are due to high-frequency thermal modes that barely contribute to the total response of the system.

\section{Experimental Setup}\label{sec:exp_setup}

Here we discuss the experimental setup used in our two experiments: the characterization of static optical nonlinearities (Section \textbf{S7}), and the measurement of optomechanical effects (Section \textbf{S8}).

\begin{figure*}[ht!]
\centering
\includegraphics[width = 8.0cm]{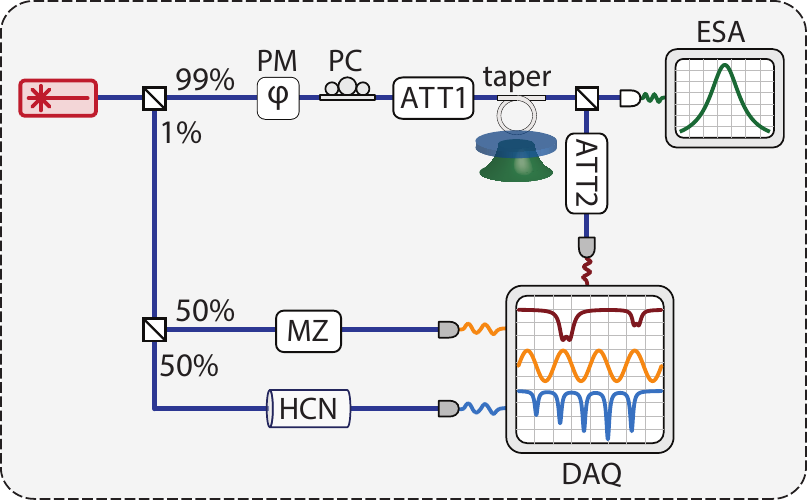}
\caption{Experimental setup: PM (Phase Modulator), ATT (Attenuator), MZ (Mach-Zehnder), HCN (HCN gas cell), PC (Polarization Controller), DAQ (Data AcQuisition system) and ESA (Electrical Spectrum Analyzer). The gray photodetectors have slow responses (response cut-off at \SI{10}{MHz}), while the white one has a fast response (cut-off at \SI{800}{MHz}).}
\label{supp:fig3}
\end{figure*}

Light emitted from a tunable laser source is sent into a (99/1) beam splitter. The 99\% output is directed to the main measurement setup, while the remaining power is used in an auxiliary frequency calibration setup, shown in Fig.~\ref{supp:fig3}, composed of a Mach-Zehnder interferometer and a referenced HCN gas cell. In the main measurement setup, light passes through a phase modulator (PM), used in the calibration of the optomechanical coupling rate; a polarization controller (PC) and a variable attenuator (ATT1) used respectively to control the polarization and intensity of the light that will be coupled to the cavity. 

Light is coupled in and out the microdisk resonator through a tapered fiber loop. The loop is supported by a carefully designed \textit{parking lot} surrounding the cavity, allowing us to probe the microdisks response without ever touching it, which would degrade both optical and mechanical quality factors. The tapered fiber output is split between two photodetectors: a slow one, linked to the DAQ (Data AcQuisition system) and used to probe the device's static optical response; and a fast one, linked to the ESA (Electrical Spectrum Analyzer), for the measurement of high frequency fluctuations of the optical signal related to the brownian mechanical motion in the optomechanical resonator. The output variable attenuator, ATT2, placed right before the slow photodetector, is used to compensate attenuation changes given by ATT1, allowing us to maintain the electronic gain settings and incident power on the slow detector fixed, avoiding both issues of signal saturation and detection nonlinearities.


In the frequency calibration branch of our setup, light is evenly split between a \SI{4}{GHz} FSR Mach-Zehnder interferometer (a relative frequency reference), and the HCN cell (an absolute frequency reference), see Fig.~\ref{supp:fig3}. The transmission spectra of both references are captured with the DAQ. Its features, such as the distance between peaks in the Mach-Zehnder and the extinctions of the HCN spectrum, are used as frequency references to construct a reliable frequency axis for the cavity's transmission spectrum. 

The optical power is measured at the input and output of the tapered fiber loop, allowing us to characterize the propagation losses of the fiber taper, \SI{-3,86}{dB}, which are both due to the taper non-adiabaticity, and due to the scattering at the region coupled to the cavity. Considering the loss is approximately symmetric in the tapered fiber, that is, the losses between the coupling region and the input and output regions are approximately equal, we can estimate the incident power at the cavity region from the input power in the fiber.



\section{Fitting of the nonlinear optical response}\label{sec:fitting}
\label{sec:nlo}

Our model predicts the photothermal effects given the absorption rate is known as a function of the cavity internal energy (number of circulating photons). Multiple linear and nonlinear processes contribute to absorption in typical semiconductor platforms, most of them are hard to determine theoretically, such as surface induced linear/two-photon absorption. Given such difficulty, we experimentally characterize the absorption rate. It is important to stress that we are not interested in carefully discriminating between the contributions of different nonlinear optical effects, but are rather focused on obtaining the heating intensity as a function of the cavity internal energy. In that sense, a phenomenological interpretation of our fitting is appropriate.

We perform measurements of the optical transmission spectrum as a function of the incident power on the cavity. For each incident power, we extract data points containing the optical frequency shift, the optical loss variation, and the cavity internal energy at resonance. Lastly, we concurrently fit such data points using a polynomial model that accounts for different kinds of optical nonlinearities.

All data and scripts discussed in this section can be accessed in Ref.~\cite{zenodo_bolometric}.

\subsection{Data collection}\label{subsec:fitting_1}

Once the fiber loop is coupled to the cavity and propagation losses are calibrated, the measurements are automated. The attenuation in ATT1 is varied from \SI{28}{dB} to \SI{8}{dB} in steps of \SI{1}{dB}. For each attenuation, the laser frequency is swept \SI{5}{nm} around the optical resonances, which are centered about \SI{1565}{nm}; the transmission is then recorded and saved. The entire experiment took approximately \SI{7}{min}, or \SI{19}{s} per power level.

\subsection{Extracting lumped parameters from the optical transmission spectra}\label{subsec:fitting_2}

\begin{figure*}[ht!]
\centering
\includegraphics[width = 16.0cm]{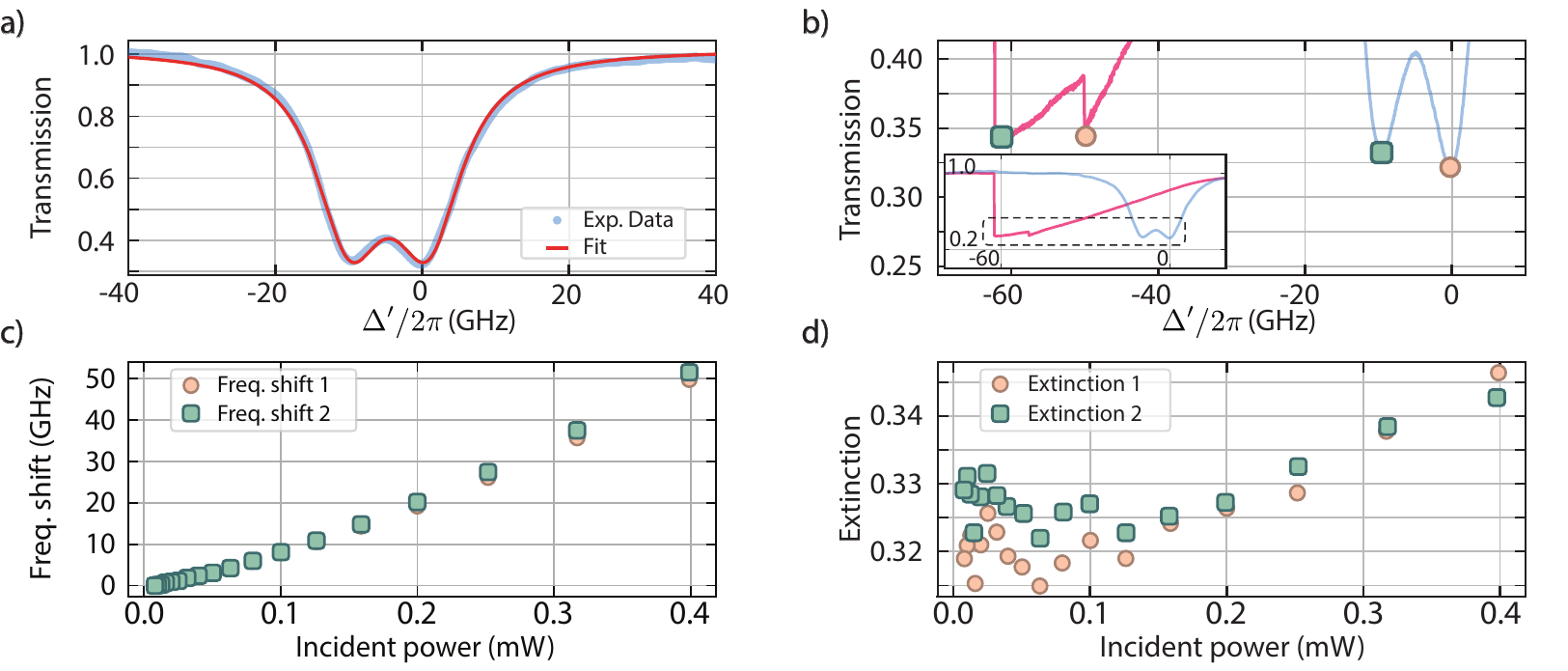}
\caption{\small{\textbf{(a)}} Fit of the cold-cavity spectrum, $P_{in}=\SI{10}{\micro\watt}$. \small{\textbf{(b)}} Comparison of the optical transmission between $P_{in}=\SI{10}{\micro\watt}$ (light red curve) and $P_{in}=\SI{399}{\micro\watt}$ (dark red curve), the markers indicate the transmission minima, the frequency shift and the extinction variation are indicated in the graph. Inset: Zoom-out of the same graph. \small{\textbf{(c)}} Measured frequency shift as a function of incident power for the first (circles) and second (squares) resonances of the doublet. \small{\textbf{(d)}} Measured extinctions as a function of incident power for the first (circles) and second (squares) resonances of the doublet.}
\label{supp:fig4}
\end{figure*}

In Fig.~\ref{supp:fig4} \textbf{a)} we display the normalized transmission spectrum of our microdisk for the second lowest incident power, it is centered around the optical doublet used in all the experiments reported in the main text. The frequency axis is the same as in Fig. \textbf{4} in the main text, its definition is $\Delta'/(2\pi) = \omega/(2\pi) - \omega_0/(2\pi) $, where $\omega$ is the absolute frequency and $\omega_0 = \SI{191.5418}{THz}$ is approximately the frequency of the higher frequency local minimum of the transmission for the low incident power data, as seen in Fig.~\ref{supp:fig4} \textbf{a)} .

In microdisks, doublets are present due to a scattering-induced coupling between degenerate clockwise and counter-clockwise propagating modes. We fit this transmission spectrum to a model given by two equal lorentzian extinction curves:
\begin{equation}
    T(\Delta) = 1-\frac{2\kappa_{i}\kappa_{e}}{4\Delta^2+\kappa^2}-\frac{2\kappa_{i}\kappa_{e}}{4(\Delta+\delta)^2+\kappa^2}.
    \label{doublet}
\end{equation}
From the fit we can obtain the cold-cavity parameters describing the doublet: the splitting between the resonances ($\delta/(2\pi)=\SI{11.3}{GHz}$), the intrinsic losses ($\kappa_{i}/(2\pi)=\SI{7.3}{GHz}$), and the extrinsic losses ($\kappa_{e}/(2\pi)=\SI{4.2}{GHz}$, related to the coupling to the tapered fiber). It is also useful to define the total loss, $\kappa =\kappa_i+\kappa_e$.

As the incident power increases, nonlinear effects change both the frequency and the linewidth of the optical doublet, giving rise to transmission spectra such as the red curve in Fig.~\ref{supp:fig4} \textbf{b)}. Results from the fit of such high power transmission spectra are unreliable, even for a simple model where nonlinear effects are polynomial functions of the internal energy. For that reason, we follow a method similar to the one used in \cite{Barclay2005NonlinearTaper} where the nonlinear effects are extracted from the transmission local minima, indicated by the circle and square markers in Fig.~\ref{supp:fig4} \textbf{b)}. For our analysis we collect the extinction (depth), $E(P_{in})$, and the frequency, $\delta\omega_{E}(P_{in})$, of the local minima for all transmission spectra. In Fig.~\ref{supp:fig4} \textbf{c)} we plot the variation of $\delta\omega_{E}(P_{in})$ while in Fig.~\ref{supp:fig4} \textbf{d)} the extinction.

In order to extract information regarding the nonlinear effects we make some assumptions for the case of an optical doublet: 1) the frequency splitting between the two resonances and the extrinsic loss, are independent of the cavity internal energy and 2) nonlinear effects are approximately equal for both resonances. Using such assumptions and the cold cavity parameters, the transmission becomes:

\begin{equation}
    T(\Delta) = 1-\frac{2(\kappa_{i}+\kappa_{nl})\kappa_{e}}{4(\Delta+\delta\omega)^2+(\kappa+\kappa_{nl})^2}-\frac{2(\kappa_{i}+\kappa_{nl})\kappa_{e}}{4(\Delta+\delta+\delta\omega)^2+(\kappa+\kappa_{nl})^2},
    \label{doublet_nonlinear}
\end{equation}
where $\kappa_{nl}$ and $\delta\omega$ are the, yet unknown, nonlinear loss and nonlinear dispersion.

Considering that the cold-cavity parameters are known from the fit and that $\delta\omega$ simply shifts the frequency axis, the extinction of the local minima, $E$, in Eq.~\ref{doublet_nonlinear} are a function only of the nonlinear loss, $E(\kappa_{nl})$. As the cavity-taper coupling remains in the undercoupled regime throughout all measurements, we can restrict our analysis to a range in which this function is bijective. Inverting it we use the extinction data in Fig.~\ref{supp:fig4} \textbf{d)} to obtain the nonlinear losses.

With the nonlinear losses we can determine the internal energy of the cavity at the extinction minima. In order to simplify our calculations we consider that the energy at the extinction minima is approximately equal the energy at the resonance points ($\Delta=0$ and $\Delta=-\delta$):
\begin{eqnarray}
    U_1 &=& U_2 = \frac{2\kappa_{e}}{(\kappa_{i}+\kappa_{nl})^2}P_{in}+\frac{2\kappa_{e}}{4\delta^2+(\kappa_{i}+\kappa_{nl})^2}P_{in}
    \label{energy}
\end{eqnarray}
As our doublet is nearly resolved this approximation is good, the difference between such points is of approximately $\Delta_E-\delta\omega~\SI{1}{GHz}$, much smaller than both the linewidth and the splitting of the resonances, rendering an error smaller than $1\%$ at the final cavity energy.

For the determination of the nonlinear dispersion we approximate the variation of the extinction minima, $\Delta_E(P_{in})$, Fig.~\ref{supp:fig4} \textbf{d)}, by the variation of the resonance frequency itself, $\delta\omega$ . Our approximation is good, rendering an error smaller than \SI{250}{MHz} for the higher incident power. This happen because of two reasons, 1) we assumed $\delta$ is fixed  and 2) the nonlinear loss variation is small when compared to $\kappa$ and $\delta$.

Since we assume that the linear losses and the nonlinear effects are equal for both modes of the doublet, data used in our analysis is obtained through an averaging process of the information collected for each of the modes in the doublet given an incident power. 

It is worth mentioning that in our determination of the nonlinear effects we took the lowest power measurement as a frequency reference, leading to two issues: 1) despite being small, the nonlinear dispersion/losses are different from zero in our reference, and 2) the extinction data for low optical powers is considerably noisy, see Fig.\ref{supp:fig4}c), such that we obtain negative values for the nonlinear loss for some of the experimental points, indicating an uncertainty at the cold-cavity losses. Both these issues can be solved with the introduction of offsets, $\delta\omega_{\text{nl,0}}$ and $\kappa_{\text{nl,0}}$ in our nonlinear model.

\subsection{Fitting of the polynomial model}\label{subsec:fitting_3}

In the fitting process we consider a polynomial model for both the nonlinear losses and the nonlinear dispersion. First we fit the nonlinear losses, and then we use this information in the fitting of the nonlinear dispersion. This method is efficient since in our model we consider only the nonlinear dispersion due to the thermo-optic effect.

In our model we consider nonlinear losses scaling both linearly and quadratically with the internal energy, $U$:

\begin{equation}
    \kappa_{nl}(U) = \kappa_{\text{nl,0}} + \kappa'_1 U + \kappa'_2 U^2.
\end{equation}
The free parameter, $\kappa_{\text{nl,0}}$, is an offset to the nonlinear losses used to correct the cold cavity intrinsic loss. The first order nonlinear loss coefficient, $\kappa'_1$, is related to two-photon absorption (TPA). It is calculated through FEM simulations where the material properties in \cite{doi:10.1063/1.3533775} are applied to the model present in \cite{Barclay2005NonlinearTaper}. The second order nonlinear loss coefficient, $\kappa'_2$, is added as a phenomenological free parameter.

First order nonlinear effects (related to TPA) are reliably negligible with respect to the quadratic effects, as seen in Fig.~\textbf{4}\textbf{d)} where the nonlinear loss displays a parabolic shape. This has also been verified by allowing the coefficient $\kappa'_1$ to vary up to one order of magnitude during the fitting process, yielding remarkably similar results to those obtained using the FEM-estimated $\kappa'_1$. 





With the knowledge of the nonlinear losses, we proceed to fit the nonlinear dispersion, given by:
\begin{equation}
    \delta\omega_{nl} = \delta\omega_{\text{nl,0}} + G^{\theta}_s R^{\theta}_s \Big(\kappa_{\text{abs}} + \eta_{nl}\big(\kappa'_1 U + B \kappa'_2 U^2\big)\Big)U.
\end{equation}
The free parameters in this step are $\delta\omega_{\text{nl,0}}$, the nonlinear dispersion offset, $\kappa_{abs}$, the absorption loss rate, and $\eta_{nl}$, the absorptive fraction of the nonlinear losses. The coefficient $G^{\theta}_s R^{\theta}_s$ is the product of the thermo-optic dispersion by the thermal resistance, for static heating sources, it is calculated with FEM simulations. We choose to neglect the change on the thermal resistance due the nonlinear losses, because simulations show that for the static case it is negligible.

As seen in Fig.~\textbf{4} \textbf{c)} and \textbf{d)}, our model fits well the experimental data, with $\kappa_{\text{nl,0}} = \SI{-1 \pm 0.1}{GHz}$, $B = \SI{1 \pm 0.1}{ns}$, $\delta\omega_{\text{nl,0}} = \SI{-5.8 \pm 0.4}{GHz}$, $\kappa_{\text{abs}} = \SI{5,05 \pm 0,006}{GHz}$ and $\eta_{nl} = \SI{0.83 \pm 0.01}{}$. As a final test of the validity of our model we "simulate" our experiment by solving the transcendental equation for the internal energy of the cavity. We use the cold cavity parameters and our nonlinear model, obtaining the transmission as a function of the laser detuning. The agreement between the experimental transmission spectra and its "simulated" counterpart, Fig.~\textbf{4} \textbf{b)}, is nearly perfect.




\section{Optical and mechanical mode simulations - Fabricated device}\label{sec:optical_id}

\subsection{Optical mode identification}\label{subsec:optical_id_1}

\begin{figure*}[ht!]
\centering
\includegraphics[width = 16.0cm]{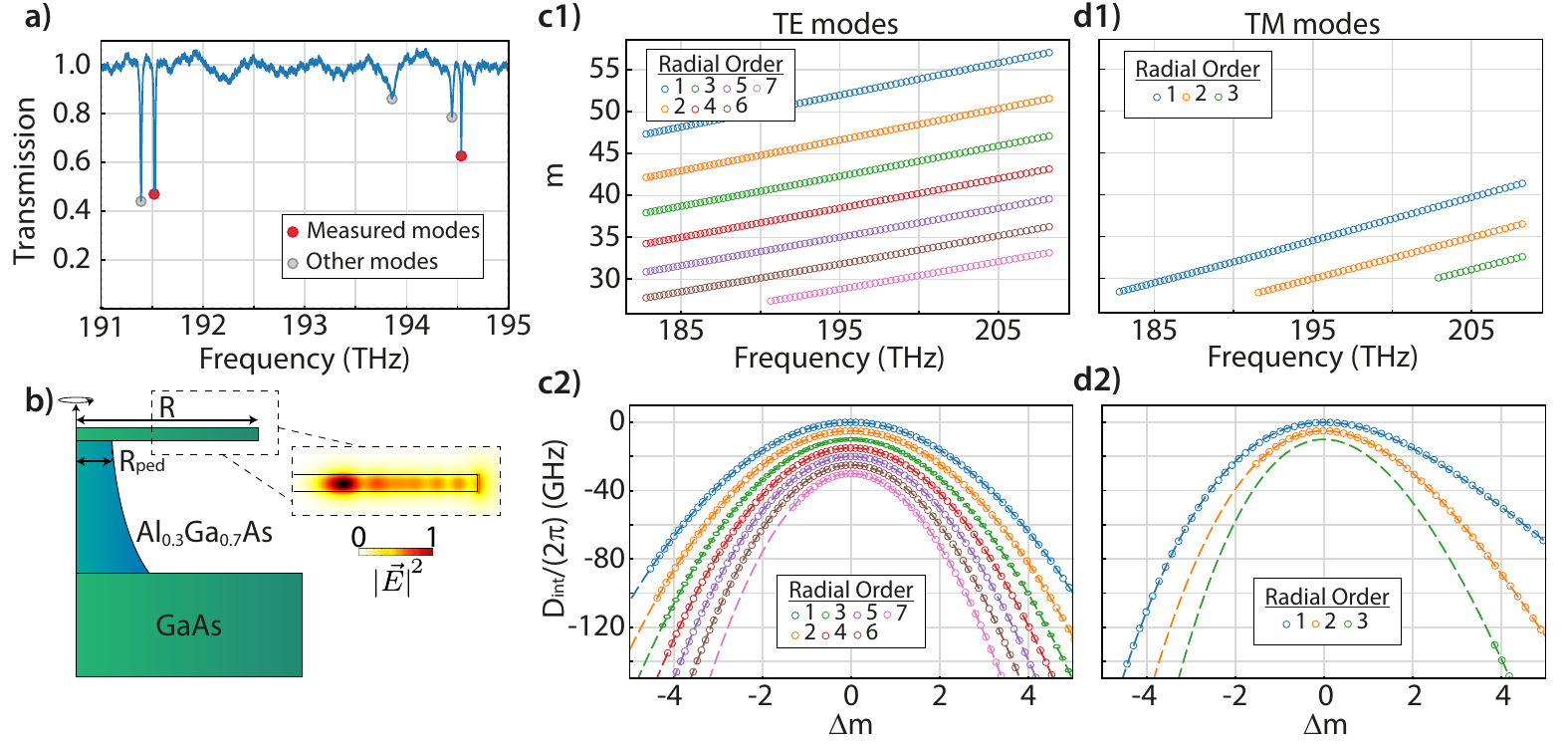}
\caption{\small{\textbf{(a)} Measured optical spectrum for the fabricated GaAs microdisk. \textbf{(b)} Schematic diagram of the device, with $R = \SI{5.1}{\micro\metre}$ and $R_\text{ped} = \SI{0.6}{\micro\metre}$. The measured $6$-th order TE optical mode is shown. \textbf{(c1)} Simulation for the optical mode's frequencies as a function of the modal number $m$ for TE modes. \textbf{(c2)} Residual dispersion for TE modes. Analogous results for the TM counterparts are shown in \textbf{(d1)} and \textbf{(d2)}.}}
\label{supp:fig1}
\end{figure*}

In order to identify the optical mode that was excited in our experiment, we analysed the optical spectrum of our microdisk as shown in Fig.~\ref{supp:fig1} \textbf{a)}. Due to the  symmetry of our device, modes can be classified through an azimuthal modal number $m$. Using a reference mode with frequency $\omega_0$ and modal number $m_0$, optical frequencies may be conveniently described as~\cite{Fujii2020DispersionGeneration}:

\begin{equation}
    \omega_{\mu}=\omega_{0}+\mu D_{1}+\frac{1}{2} \mu^{2} D_{2}+\frac{1}{6} \mu^{3} D_{3}+\ldots,
\end{equation}
where $\mu = m-m_0$, $D_{1}/(2\pi)$ is the Free Spectral Range (FSR) and the $D_{n}/(2\pi)$, $n>1$, give the FSR dependency on $\mu^{n-1}$.

For the measured optical modes FSR $\approx \SI{3}{\tera\hertz}$. We compare this value with simulations that incorporate both geometric and material dispersions for our device. Our measurements are found to be consistent with the $6$-th radial order TE optical mode shown in Fig.~\ref{supp:fig1} \textbf{b)}. In Fig.~\ref{supp:fig1} \textbf{c1)} (\textbf{d1}) the $m$ dependency with the frequency is plotted for TE (TM) optical modes between $180$ and $\SI{210}{\tera\hertz}$. For completeness, in Fig.~\ref{supp:fig1} \textbf{c2)} (\textbf{d2}) we also show the residual dispersion $D_\text{int} = \omega_\mu-\omega_0-\mu D_{1}$ for the simulated TE (TM) modes. Results are summarized in Table~\ref{supp:tab1}.

\begin{table}[H]
\centering
\begin{tabular}{c|c|c}
\hline
Mode & $D_1/(2\pi)$ (THz) & $D_2/(2\pi)$ (GHz) \\ \hline
TE$_1$ & 2.61 & -8.0 \\ \hline
TE$_2$ & 2.70 & -10.0 \\ \hline
TE$_3$ & 2.78 & -12.0 \\ \hline
TE$_4$ & 2.85 & -13.0 \\ \hline
TE$_5$ & 2.92 & -16.0 \\ \hline
TE$_6$ & 2.99 & -18.0 \\ \hline
TE$_7$ & 3.07 & -22.0 \\ \hline
TM$_1$ & 1.95 & -9.0 \\ \hline
TM$_2$ & 2.05 & -15.0 \\ \hline
TM$_3$ & 2.16 & -21.0
\end{tabular}
\caption{$D_1$ and $D_2$ for each of the optical modes in Figs.~\ref{supp:fig1} \textbf{(c2)} and \textbf{(d2)}.}
\label{supp:tab1}
\end{table}

\subsection{Mechanical anisotropy}\label{subsec:optical_id_2}

We incorporate the known mechanical anisotropy of GaAs in a 3D numerical model of the microdisk. This is done by correctly prescribing values for the stiffness tensor coefficients used in our simulations: (using Voigt notation) $c_{11} = \SI{119}{\giga\pascal}$, $c_{12} = \SI{53.4}{\giga\pascal}$, and $c_{44} = \SI{59.6}{\giga\pascal}$ \cite{Balram2014MovingResonators}. The symmetries of the system allow for correct calculations of the acoustic eigenmodes by considering only 1/4 of the full structure. All overlap integrals between optical, thermal, and mechanical modes were carefully evaluated by considering isotropic models for the thermal and optical responses. The former are calculated from 2D axisymmetric simulations for the device, which are then extruded into fully 3D fields. Only by taking in account the GaAs anisotropy we were able to reach an agreement between the simulation and the experimental data for the single photon optomechanical coupling, $g_0/2\pi=\SI{29}{kHz}$.

\section{Optomechanical response}\label{sec:om_response}

In this section we discuss in detail how the experimental data on the optomechanical response was collected, analyzed and compared with the results expected from our model. 

\begin{figure*}[h!]
\centering
\includegraphics[width = 16.0cm]{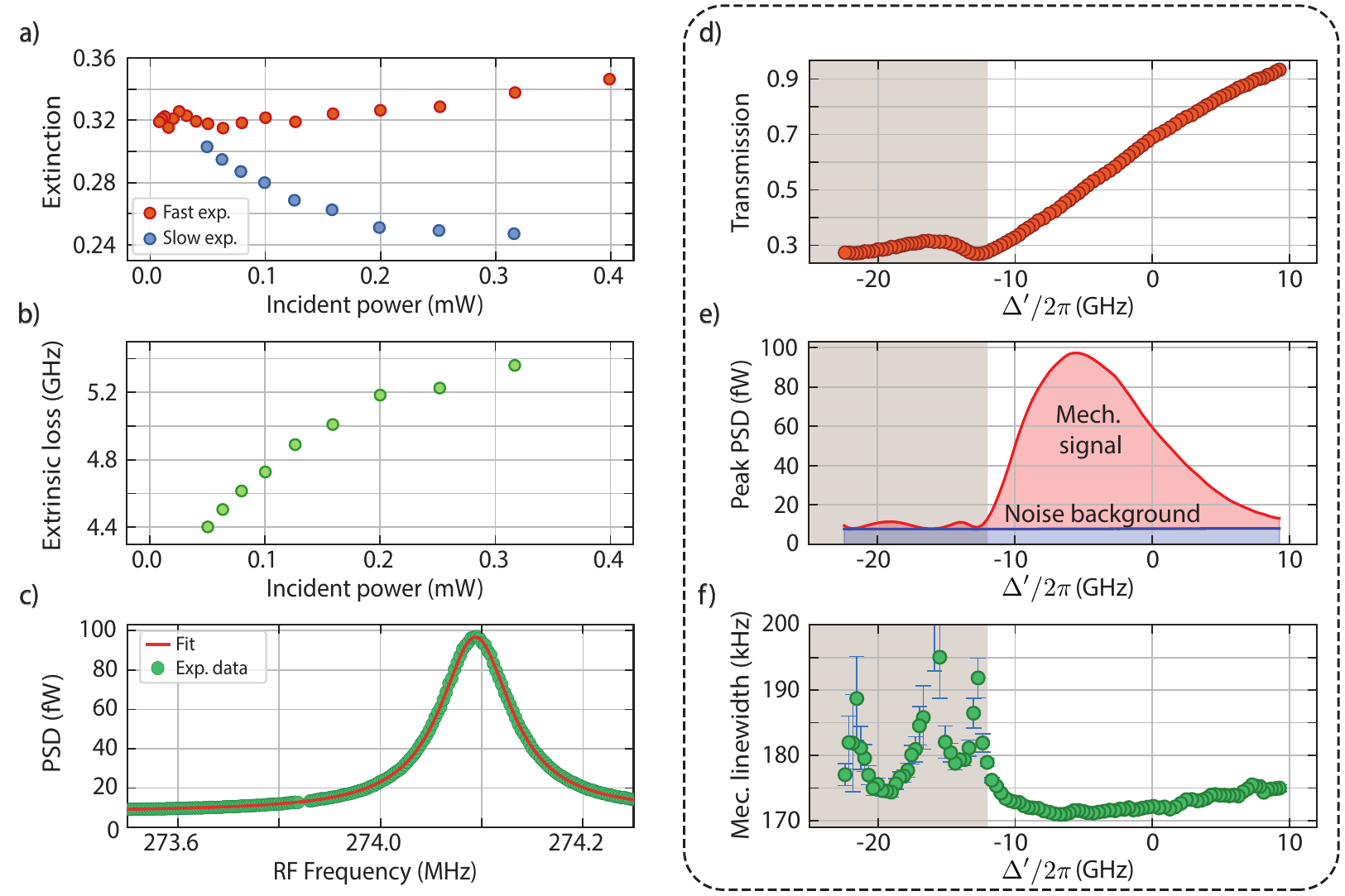}
    \caption{\small{\textbf{(a)} Extinction of the optical transmission as a function of the incident power, for both the optical nonlinear characterization, or fast experiment (red dots); and for the optomechanical characterization, or slow experiment (blue dots). \textbf{(b)} Extrinsic loss during the slow experiment. \textbf{(c)} PSD with fit. \textbf{(d)} Optical transmission collected during the measurement of the PSD map. \textbf{(e)} PSD peak as a function of the optical detuning. \textbf{(f)} Mechanical linewidth variation as a function of the optical detuning. In graphs\textbf{(c)},\textbf{(d)},\textbf{(e)} and \textbf{(f)} the incident power is \SI{126}{\micro\watt}. The brown shade in the graphs indicates the higher optomechanical transduction.}} 
\label{supp:fig5}
\end{figure*}

\subsection{Data collection}\label{subsec:om_response_1}

For greater precision, all optomechanical measurements were performed by fine-tuning the laser frequency with a piezoelectric actuator. This choice limits the measurable optical frequency span to $\SI{30}{\GHz}$ around a center frequency, which is chosen to render the peak modification on the mechanical linewidth within the measured range. Furthermore, the thermally-induced optical bistability
restricts our experiments to the blue side of the optical resonance for the higher incident powers used in this work. 

In this experiment we use a fast photodetector, with a \SI{800}{MHz} bandwidth, to measure the mechanical fluctuations induced in the optical signal. The electrical signal of the photodetector is monitored with an Electrical Spectrum Analizer (ESA), which returns the Power Spectral Density (PSD) of its input signal. Fig.\textbf{4} \textbf{f)} in the main text is an example of a PSD where the fluctuations due to the fundamental breathing mode of our microdisk clearly rise above the detector's noise floor. As the optical fluctuations are very weak, each ESA measurement must be integrated for some seconds.


For each attenuation a regular, static, optical transmission spectrum is measured using the standard laser frequency sweep -- as described in section \ref{sec:nlo} --, and a PSD map varying the laser frequency with the piezoelectric actuator, Fig.\textbf{4}\textbf{e)}. A PSD map is composed of 100 PSD taken at equally spaced laser frequencies inside the \SI{30}{GHz} piezo range. The laser spends approximately \SI{3}{s} in each frequency, in such a way that building a PSD map for one attenuation takes approximately \SI{5}{min} and \SI{30}{s}. The attenuation is varied from \SI{20}{dB} to \SI{12}{dB} in steps of \SI{1}{dB}. The entire experiment takes around \SI{1}{h}. 


As this experiment is considerably longer than the previous one, the tapered fiber position drifts moderately with respect to the microdisk, changing the extrinsic loss of our optical resonances. This is clearly seen in Fig.~\ref{supp:fig5} \textbf{a)} where the extinctions of the optical spectra taken in the optomechanical characterization (slow experiment) and in the optical nonlinear characterization (fast experiment) are compared. For lower incident powers the extinctions of both are similar, but as the $P_{in}$ increases the extinction of the slow experiment becomes smaller than the one in the fast experiment. As the slow experiment is performed just after the fast one, this indicates that the taper position slowly drifts closer to the cavity as a function of time, increasing its optical coupling to the cavity.

We use the extinction information to correct for the extrinsic loss. Since the static optical transmission is recorded for each of the input powers in the fast experiment, we are able to recover its resonance frequency shift and compare it to the data obtained in the slow experiment. This gives us direct knowledge of the energy (at the extinction minima) in the resonator during the fast experiment. We use this information to determine the nonlinear loss, in such a way that the extinction becomes now a function only of the extrinsic loss. In a similar way to what was previously done for the determination of the nonlinear loss,  we derive a numerical bijective relation between the extrinsic loss and the extinction, which we can invert to obtain the extrinsic loss, as shown in Fig.~\ref{supp:fig5}\textbf{b)}. This estimate of the extrinsic loss during the optomechanical characterization is applied to our model for the photothermal effects. 

\subsection{Mechanical linewidth variation}\label{subsec:om_response_2}


Fitting the PSD to a model taking into account the mechanical susceptibility and a constant noise background, as in Fig.~\ref{supp:fig5} \textbf{c)}, we can determine the mechanical frequency and linewidth as a function of the laser frequency. 

In order to study photothermal effects we analyzed the mechanical linewidth variation. In Fig.~\ref{supp:fig5} f) two distinct behaviors are verified. For $\Delta'$ between \SI{10}{GHz} and \SI{-12}{GHz} (clear region), the mechanical linewidth varies smoothly and presents very small uncertainty bars, while for $\Delta'<\SI{-12}{GHz}$ (shaded region), uncertainty increases considerably. As seen in Fig.~\ref{supp:fig5} d), the clear region is found to match the blue detuned lateral of the optical resonance, where the high optomechanical transduction, shown in Fig.~\ref{supp:fig5} e), leads to a large signal to noise ratio (SNR) in the measurement of the mechanical linewidth. Due to its larger precision, we focus our analysis on the clear region. The shaded region is related to the transmission minima, where the optomechanical transduction approaches zero, leading to a low SNR in the PSD.




Feeding our model for the photothermal effects with the optical nonlinearities, and with the extrinsic loss estimate, we can predict effects such as the mechanical linewidth variation. Fig.\textbf{4}\textbf{g)} and \textbf{h)} of the main text show our predictions as a function of both the laser detuning and the incident power at the cavity, matching very well with the experimental data. The value of the parameters fed to our model are listed in Tab.\ref{tab:parameters} as well as a brief explanation of how they were obtained. Here we must stress that, differently from other works, we do not fit the optomechanical data, instead we predict it independently using FEM simulations and the characterization of the optical nonlinearities in the resonator. As discussed in the main text, the use of multiple thermal modes is crucial to the level of accuracy obtained here.

The geometrical dimensions ($R = \SI{5.1\pm0.1}{\micro\metre}$ and $R_\text{ped} = \SI{0.6\pm0.1}{\micro\metre}$) of our device are roughly determined from optical microscopy data. The value for the radius used in all simulations is fine-tuned matching the mechanical frequency from FEM simulations with the experimental mechanical frequency.

\subsection{Mechanical frequency shift}\label{subsec:om_response_3}

\begin{figure*}[ht!]
\centering
\includegraphics[width = 16.0cm]{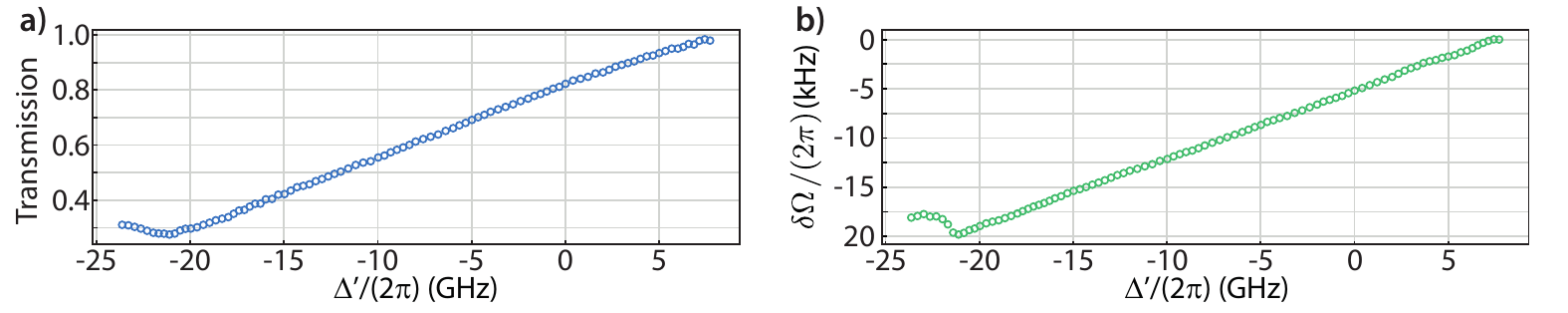}
\caption{\small{\textbf{(a)} Measured optical spectrum for the fabricated GaAs microdisk under an excitation of $\SI{126}{\micro\watt}$. \textbf{(b)} Mechanical frequency modification for the same incident power.}}
\label{supp:fig2}
\end{figure*}

As mentioned in the main text, the mechanical frequency modification is dominated by thermalization of the cavity. Qualitatively, the heat generated by the optical absorption softens the GaAs membrane, which in turn downshifts the mechanical frequencies. This competes with the positive mechanical frequency shift predicted due to optomechanical backaction (for blue detunings), as shown in Fig.~\textbf{3} \textbf{(a)} of the main text. Plots for the measured transmission spectrum and mechanical frequency shift $\delta \Omega$ are found in Figs.~\ref{supp:fig2} \textbf{(a)} and \textbf{(b)}, where one readily observes that the trend followed by $\delta \Omega$ strikingly contrasts with the one followed by $\delta \Gamma$ (Fig.~\textbf{4} \textbf{(g)}, main text). In fact, the transmission and $\delta \Omega$ display similar patterns. This is a direct consequence of the increase in the circulating power when the laser approaches the optical resonance, increasing the overall temperature in the resonator.

\vspace{10pt}
\textbf{Dataset}:FEM and scripts files for generating each figure are available at Ref.~\cite{zenodo_bolometric}.


%